\documentclass[a4paper,11pt]{article}
\pdfoutput=1 

\usepackage{jcappub} 

\usepackage[T1]{fontenc} 
\usepackage{siunitx}
\usepackage[compat=1.1.0]{tikz-feynman}
\usepackage{comment}
\usepackage{slashed}
\usepackage{physics}
\usepackage{dsfont}
\usepackage{subcaption}
\usepackage{tensor}
\usepackage{bbold}
\usepackage[normalem]{ulem}
\usepackage{booktabs}

\newcommand\MSbar{\ensuremath{\overline{\mathrm{MS}}}}
\newcommand\aP{|\mathbf{P}|}
\newcommand\ak{|\mathbf{k}|}

\newcommand\al{|\mathbf{l}|}
\newcommand\aq{|\mathbf{q}|}
\newcommand\ap{|\mathbf{p}|}

\newcommand{\alphaEM}{\alpha_{\rm em}}
\newcommand{\HTLeq}{\overset{\mathrm{HTL}}{\approx}}
\renewcommand{\d}{\mathrm{d}}

\newcommand{\fFermi}{f_F}

\title{\boldmath Towards a precision calculation of 
$N_{\rm eff}$ in the Standard Model III: Improved estimate of NLO contributions to the collision integral}

\author[a]{Marco~Drewes,}
\author[a]{Yannis~Georis,}
\author[b]{Michael~Klasen,}
\author[b]{Luca~Paolo~Wiggering}
\author[c]{and Yvonne~Y.~Y.~Wong}

\affiliation[a]{Centre for Cosmology, Particle Physics and Phenomenology, Universit\'{e} catholique de Louvain, Louvain-la-Neuve B-1348, Belgium}
\affiliation[b]{Institut f\"ur Theoretische Physik, Universit\"at M\"unster, Wilhelm-Klemm-Stra{\ss}e 9, D-48149 M\"unster, Germany}
\affiliation[c]{Sydney Consortium for Particle Physics and Cosmology, School of Physics, The University of New South Wales, Sydney NSW 2052, Australia}

\emailAdd{marco.drewes@uclouvain.be, yannis.georis@uclouvain.be, michael.klasen@uni-muenster.de, luca.wiggering@uni-muenster.de, yvonne.y.wong@unsw.edu.au}

\abstract{We compute the dominant 
QED correction to the neutrino-electron interaction rate in the vicinity of neutrino decoupling in the early universe, and estimate its impact on the effective number of neutrino species $N_{\rm eff}$ in cosmic microwave background anisotropy observations. We find that the correction to the interaction rate is at the sub-percent level, consistent with a recent estimate by Jackson and Laine.  Relative to that work we include the electron mass in our computations, but restrict our analysis to the enhanced $t$-channel contributions. The fractional change in $N_{\rm eff}^{\rm SM}$ due to the rate correction is of order $10^{-5}$ or below, i.e., about a factor of 30 smaller than that recently claimed by Cielo {\it et al.}, and below the nominal computational uncertainties of the current benchmark value of $N_{\rm eff}^{\rm SM} = 3.0440 \pm 0.0002$.  We therefore conclude that aforementioned number remains to be the state-of-the-art benchmark for $N_{\rm eff}^{\rm SM}$ in the standard model of particle physics. 
}

\begin{document}
\begin{flushright}
	{\large \tt CPPC-2024-01, MS-TP-24-06}
\end{flushright}

\maketitle
\flushbottom

\section{Introduction}

The effective number of neutrinos, $N_{\rm eff}$, is an important parameter in standard hot big bang cosmology~\cite{Steigman:1977kc}.  Defined as the energy density residing in free-streaming, ultra-relativistic particle species relative to the photon energy density in the  post-neutrino decoupling early universe (i.e., at temperature $T \lesssim 1$~MeV), the primary role of $N_{\rm eff}$ is to fix the universal expansion rate at $T \lesssim 1$~MeV up to the end of the radiation-domination epoch.
Its observable consequences are many---from setting the primordial light element abundances, to influencing the correlation statistics of the cosmic microwave background (CMB) anisotropies and the large-scale matter distribution.  Coupled with the fact that many beyond-the-standard-model scenarios predict $N_{\rm eff}$-like effects~(e.g.,  light sterile neutrinos~\cite{Barbieri:1990vx,Abdullahi:2022jlv}, axions~\cite{DiLuzio:2022gsc,DEramo:2021lgb},
gravitational waves~\cite{Caprini:2018mtu}, hidden sectors~\cite{Aboubrahim:2022gjb,Agrawal:2021dbo}, etc.), pinning down its value both theoretically and observationally has enjoyed an unwavering interest for over four decades~\cite{Steigman:1977kc,Aghanim:2018eyx}.

\begin{figure}[t]
    \hspace{.5cm}
    \includegraphics[width=.85\textwidth]{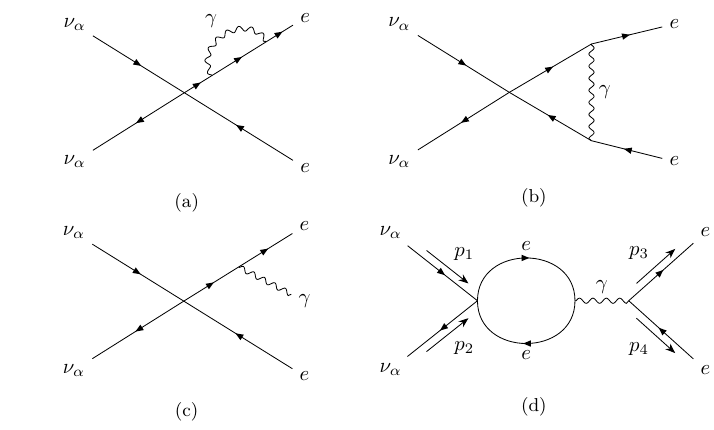}
    \caption{The four qualitatively different QED corrections to the neutrino interaction rate corresponding to (a)~modification of the electron dispersion relation, (b)~virtual photon exchange, (c)~thermal photon emission and absorption, and (d) corrections with a closed fermion loop. All diagrams are schematic in that all time directions are possible.}
    \label{fig:weakRates2}
\end{figure}

From the theoretical perspective, the expected value of $N_{\rm eff}$ in the context of the standard model (SM) of particle physics is 3 (for three generations), plus percent-level corrections due to residual energy transfer between the quantum electrodynamic (QED) plasma and the neutrino sector during neutrino decoupling~\cite{Dodelson:1992km,Hannestad:1995rs,Dolgov:1997mb,Dolgov:1998sf,Esposito:2000hi,Froustey:2019owm} as well as deviations of the QED plasma itself from an ideal gas~\cite{Dicus:1982bz,Heckler:1994tv,Fornengo:1997wa,Lopez:1998vk,Mangano:2001iu,Bennett:2019ewm}.
Historical estimates of these corrections have ranged from $0.011$ to $0.052$~\cite{Dolgov:1997mb,Mangano:2005cc,Birrell:2014uka,Grohs2016,deSalas:2016ztq}.
Detailed modelling in recent years~\cite{Gariazzo:2019gyi,Akita:2020szl,Froustey:2020mcq,Bennett:2020zkv}, however, have drastically reduced the spread. Notably, two of us (Drewes and Wong) and collaborators reported in~\cite{Bennett:2020zkv} a prediction of $N_{\rm eff}^{\rm SM} = 3.0440 \pm 0.0002$ from a fully momentum-dependent precision transport calculation that accounted for (i)~neutrino oscillations, (ii)~finite-temperature corrections to the QED equation of state (EoS) to ${\cal O}(e^3)$, and (iii) a first estimate of finite-temperature corrections of type~(a) to the weak interaction rates depicted in figure~\ref{fig:weakRates2} (see also table~\ref{tab:Split}); the error bars are due mainly to numerical resolution and experimental uncertainties in the input neutrino mixing angles. 
More remarkably still, this result is in perfect agreement with the independent calculations of~\cite{Akita:2020szl,Froustey:2020mcq} modelling the same physics---to five significant digits in the central value and with comparable error estimates.
The precision computation of $N_{\rm eff}^{\rm SM}$ appears therefore to have reached convergence, at least in a limited sense.

\begin{table}[t]
\centering
\begin{tabular}{lc}
\toprule
Standard-model corrections to $N_{\rm eff}^{\rm SM}$ & Leading-digit contribution \\
\midrule
$m_e/T_d$ correction& $+0.04$ \\
$\mathcal{O}(e^2)$ FTQED correction to the QED EoS& $+0.01$\\
Non-instantaneous decoupling+spectral distortion & $-0.006$\\
$\mathcal{O}(e^3)$ FTQED correction to the QED EoS& $-0.001$\\
Flavour oscillations & $+0.0005$\\
Type (a) FTQED corrections to the weak rates & $\lesssim 10^{-4}$\\
\bottomrule
\toprule
Sources of uncertainty & \\
\midrule
Numerical solution by {\tt FortEPiaNO} & $\pm 0.0001$ \\
Input solar neutrino mixing angle $\theta_{12}$ &  $\pm 0.0001$ \\
\bottomrule
\end{tabular}
\caption{Leading-digit contributions from various SM corrections, in order of importance, thus far accounted that make up the final $N_{\rm eff}^{\rm SM}-3=0.0440 \pm 0.0002$~\cite{Bennett:2020zkv,Froustey:2020mcq}.  Table adapted from~\cite{Bennett:2020zkv}.}
\label{tab:Split}
\end{table}


There are nonetheless reasons to be cautious.  For one, while the general expectation is that the physics summarised in table~\ref{tab:Split} dominate corrections to $N_{\rm eff}^{\rm SM}$, as yet missing is a systematic study of possible higher-order effects that may be at least as important.  Indeed, in estimating only next-to-leading-order (NLO) effects due to diagram (a) to the weak rates rather than the full range of corrections displayed in figure~\ref{fig:weakRates2}, one could argue that the computations of~\cite{Akita:2020szl,Froustey:2020mcq,Bennett:2020zkv} are, at least conceptually, incomplete.  Two recent works have taken a first step towards filling this gap: 
\begin{itemize}
\item Cielo {\it et al.}~\cite{Cielo:2023bqp} took the finite-temperature rate corrections for $e^+ e^- \to \nu_\alpha \bar{\nu}_\alpha$ from~\cite{Esposito:2003wv}, computed originally in the context of energy loss in a stellar plasma, and appealed to detailed balance to estimate the corrections to collision integrals. The claimed effect of this correction on $N_{\rm eff}^{\rm SM}$ is quite substantial---at the $\sim 0.001$ level.
We have reservations about this result:
Aside from mapping rate corrections from an incompatible energy regime (${\cal O}(1)$~keV $\ll m_e$ in a stellar plasma versus ${\cal O}(1)$~MeV $> m_e$ in the early universe), the analysis of~\cite{Cielo:2023bqp} also neglected (i)~corrections due to diagram~(d) in figure~\ref{fig:weakRates2} (the ``closed fermion loop''), (ii)~corrections to elastic scattering reactions like $\nu_\alpha e \to \nu_\alpha e$, as well as (iii)~Pauli blocking effects of the final-state neutrinos. Of particular note is that the neglected closed fermion loop diagram~(d) contains a $t$-channel enhancement, which should, at least na\"{i}vely, dominate the weak rate corrections.

\item The more recent work of Jackson and Laine~\cite{Jackson:2023zkl} considered all diagrams in figure~\ref{fig:weakRates2} in a first-principles calculation using the {\it imaginary-time} formalism of thermal field theory   which accounts for both vacuum and finite-temperature corrections,%
\footnote{We note that computations of the vacuum corrections to the neutrino-electron interaction due to diagrams (a)--(d) already appeared previously in reference~\cite{Tomalak:2019ibg}.}
along with an estimate of hadronic corrections to diagram (d) following~\cite{Hill:2019xqk}.  Because the imaginary-time formalism applies only to systems in thermal equilibrium, the computation of~\cite{Jackson:2023zkl} effectively neglected non-equilibrium neutrino phase space distributions.  Additionally, 
the authors assumed a negligible electron mass by setting $m_e=0$, which is not necessitated by the formalism. 
The final result, presented as a set of  corrections to the leading-order (LO) neutrino damping rate, confirms the expected $t$-channel enhancement in diagram~(d). Nevertheless, these corrections are minute---of order 0.2 to 0.3\%.
While the authors of~\cite{Jackson:2023zkl} did not report the corresponding change in $N_{\rm eff}^{\rm SM}$, it is clear that corrections of this magnitude cannot effect a shift in $N_{\rm eff}$ as sizeable as that claimed in~\cite{Cielo:2023bqp}.\looseness=-1
\end{itemize}

The purpose of the present work is to clarify whether or not QED corrections to the neutrino interaction rate can alter the standard-model $N_{\rm eff}^{\rm SM}$ at the $\sim  0.001$ level as claimed in \cite{Cielo:2023bqp}.  
While this correction is small relative to the anticipated  sensitivity of the next-generation CMB-S4 program to $N_{\rm eff}$,  $\sigma(N_{\rm eff}) \simeq 0.02-0.03$~\cite{CMB-S4:2016ple}, an accurate theoretical prediction for $N_{\rm eff}^{\rm SM}$ at per-mille level accuracy is nonetheless desirable to justify neglecting the theoretical uncertainty in cosmological parameter inference.  In this regard, our work shares a common motivation with Jackson and Laine~\cite{Jackson:2023zkl}.  However, our work also differs from~\cite{Jackson:2023zkl} in three important ways: 
\begin{enumerate}
\item Our first-principles computation of the QED corrections uses the so-called {\it real-time} formalism and focuses on corrections of type~(d) which contain a $t$-channel enhancement.
Computation of the non-enhanced (and hence sub-dominant) diagrams (a)--(c) is postponed to a follow-up work.  Like the imaginary-time formalism, the real-time formalism also automatically takes care of both vacuum and finite-temperature corrections.
While in equilibrium situations the two formalisms are exactly equivalent~\cite{Landsman:1986uw}, the latter has the advantage that it can easily be generalised to non-equilibrium settings~\cite{Chou:1984es,Berges:2004yj}.
Thus, although we shall restrict the present analysis to the same equilibrium conditions as in~\cite{Jackson:2023zkl} and hence  provide an independent partial cross-check of their results, our calculation will also pave the way for incorporating NLO effects in neutrino decoupling codes such as {\tt FortEPianNO}~\cite{Gariazzo:2019gyi} to deliver a final verdict on $N_{\rm eff}^{\rm SM}$.

\item We retain a finite electron mass $m_e$ in our calculation, which represents a departure from the $m_e=0$ approximation made in~\cite{Jackson:2023zkl}. Neutrino decoupling occurs at temperatures $T \sim 1$~MeV, in whose vicinity the setting of $m_e=0$ may not be well justified.  We shall examine the validity of the approximation and its impact on $N_{\rm eff}^{\rm SM}$.

\item Finally, we use our NLO results to estimate the corresponding change in $N_{\rm eff}^{\rm SM}$.  This estimate was missing in~\cite{Jackson:2023zkl}.

\end{enumerate}

The rest of the article is organised as follows.  In section~\ref{sec:prelim} we describe the physical system and sketch how $N_{\rm eff}^{\rm SM}$ is computed.  Section~\ref{sec:calc} outlines the computation of QED corrections to the neutrino damping rate due to diagram~(d) and presents our numerical estimates of these rate corrections.  
The shift in $N_{\rm eff}^{\rm SM}$ due to these corrections is presented in section~\ref{sec:Neffestimates}, and section~\ref{sec:conclusions} contains our conclusions.  Five appendices provide details on the bosonic and fermionic propagators at finite temperatures, explicit expressions for certain thermal integrals, derivation of the 1PI-resummed photon propagator, parameterisation of the collision integrals, and functions appearing in the continuity equation.


\section{Preliminaries}
\label{sec:prelim}

The SM effective number of neutrinos $N_{\rm eff}^{\rm SM}$ is defined via the ratio of the energy density contained in three generations of neutrinos and antineutrinos, $\rho_{\nu}$, to the photon energy density $\rho_\gamma$ in the limit $T/m_e \to0$,
\begin{equation}
\label{NeffFermief}
\left. \frac{\rho_\nu}{\rho_\gamma} \right|_{T/m_{e} \to 0} \equiv \frac{7}{8} \Big(\frac{4}{11}\Big)^{4/3} N_{\rm eff}^{\rm SM}.
\end{equation}
Here, $T$ is the photon temperature, $m_e$ the electron mass, and the limit $T/m_e \to 0$ is understood to apply only to $T \gg m_i$, where $m_i$ is the largest neutrino mass.  

To compute the precise value of $N_{\rm eff}^{\rm SM}$ requires that we track the evolution of $\rho_\nu$ and $\rho_\gamma$ simultaneously across the neutrino decoupling epoch, i.e., $T \sim {\cal O}(10) \to {\cal O}(0.01)$~MeV.  Assuming that at these temperatures the photons are held in a state of thermodynamic equilibrium together with the electrons/positrons--- collectively referred to as the ``QED plasma''---this is typically achieved by solving two sets of evolution equations: (i) a continuity equation that follows the total energy density of the universe, and (ii) some form of generalised Boltzmann equations---often referred to as the quantum kinetic equations (QKEs)---which describe the non-equilibrium dynamics in the neutrino sector across its decoupling.

\paragraph{Continuity equation}
In a Friedmann-Lema\^{\i}tre-Robertson-Walker (FLRW) universe, the continuity equation is given by
\begin{equation}
\label{eq:continuity}
\frac{{\rm d} \rho_{\rm tot}}{{\rm d} t}  + 3 H (\rho_{\rm tot} + P_{\rm tot}) = 0,
\end{equation}
where $\rho_{\rm tot}$ and $P_{\rm tot}$ are, respectively, the total  energy density and pressure, $H\equiv (1/a) ({\rm d} a/{\rm d} t)$ the Hubble expansion rate, and $a$ is the scale factor. For the physical system at hand, $\rho_{\rm tot} \equiv \rho_{\rm QED} + \rho_\nu$ 
and $P_{\rm tot} \equiv P_{\rm QED}+P_\nu$, where $\rho_{\rm QED} \equiv \rho_\gamma + \rho_e$ subsumes the photon and the electron/positron energy densities, and similarly for $P_{\rm QED}$.  The assumption of thermodynamic equilibrium in the QED sector in the time frame of interest means that the standard thermodynamic relation $\rho_{\rm QED} = -P_{\rm QED}+ T \, (\partial /\partial T) P_{\rm QED}$ applies. Then, the finite-temperature corrections to the QED equation of state summarised in table~\ref{tab:Split} simply refer to corrections to the QED partition function $Z_{\rm QED}$ and hence $P_{\rm QED} = (T/V) \ln Z_{\rm QED}$ that alter $\rho_{\rm QED} + P_{\rm QED}= T \, (\partial /\partial T) P_{\rm QED}$ from its ideal-gas expectation. Corrections to $Z_{\rm QED}$ are known to ${\cal O}(e^3)$ for arbitrary $m_e$ and chemical potential $\mu$~\cite{Kapusta:2006pm} and to ${\cal O}(e^5)$ for $m_e = \mu=0$~\cite{Coriano:1994re,Parwani:1994xi}.  Their effects on $N_{\rm eff}^{\rm SM}$ have been estimated in references~\cite{Heckler:1994tv,Mangano:2001iu,Bennett:2019ewm} to ${\cal O}(e^4)$.

\paragraph{Quantum kinetic equations}  State-of-the-art neutrino decoupling calculations employ the QKEs to track simultaneously the effects of 
in-medium neutrino flavour oscillations
and particle scatterings on the one-particle reduced density matrix of the neutrino ensemble, $\varrho=\varrho(p,t)$.
Schematically, the QKEs take the form~\cite{Sigl:1993ctk}
\begin{equation}
\label{eq:qke}
\partial_t \varrho - p H \partial_p \varrho = - {\rm i} [\mathbb{H},\varrho] + {\cal I}[\varrho],
\end{equation}
where $\mathbb{H}=\mathbb{H}(p,t)= \mathbb{H}_{\rm vac} + \mathbb{V}$ is the effective Hamiltonian incorporating vacuum flavour oscillations $\mathbb{H}_{\rm vac}$ and in-medium corrections from forward scattering $\mathbb{V}$,%
\footnote{Depending on context, these modifications to the in-medium quasiparticle dispersion relations are variously known as thermal masses, matter potentials, or refractive indices.}
and 
\begin{equation}
\label{eq:collisionintegral}
{\cal I}[\varrho]   =\frac{1}{2}\Big((\mathbb{1}-\varrho) \mathbb{\Gamma}^< - \varrho \mathbb{\Gamma}^> \Big) + {\rm h.c.}
\end{equation}
is the collision integral encapsulating the non-unitary gains ($\mathbb{\Gamma}^<=\mathbb{\Gamma}^<(p,t)$) and losses ($\mathbb{\Gamma}^>=\mathbb{\Gamma}^>(p,t)$) of $\varrho$.  In the context of a precision calculation of $N_{\rm eff}^{\rm SM}$, the quantities $\varrho$, $\mathbb{H}$, and $\mathbb{\Gamma}^\gtrless$ are understood to be $3 \times 3$ hermitian matrices in flavour space, with the diagonal entries of $\varrho$ corresponding to the occupation numbers $f_\alpha(p,t) \equiv \{\varrho (p,t)\}_{\alpha \alpha}$, for $\alpha = e, \mu,\tau$.%
\footnote{Strictly speaking, an occupation number refers to a diagonal element of $\varrho$ in 
the basis in which $\mathbb{H}$ is diagonal, i.e., in the mass basis, rather than the interaction basis. This distinction is however unimportant for the present study, and we follow the common practice of calling, e.g., $\{\varrho (p,t)\}_{ee}$ the electron neutrino occupation number, etc.}
We assume a $CP$-symmetric universe, which is well justified if any asymmetry in the lepton sector mirrors the baryon asymmetry in the observable universe~\cite{Canetti:2012zc}.  In practice this means it suffices to follow only one set of QKEs for the neutrinos; the antineutrinos evolve in the same way.

For the problem at hand, the gain and loss terms $\mathbb{\Gamma}^\gtrless$ incorporate in principle all weak scattering processes wherein at least one neutrino appears in either the initial or final state. In the temperature window of interest, however, it suffices to consider only $2 \to 2$ processes involving (i) two neutrinos and two electrons any way distributed in the initial and final states (labelled ``$\nu e$''), and (ii) neutrino-neutrino scattering (``$\nu \nu$''). The leading-order $\mathbb{\Gamma}^\gtrless$ for these processes are well known, and take the form of two-dimensional momentum integrals,
\begin{equation}
\begin{aligned}
\label{eq:twoD}
\mathbb{\Gamma}^\gtrless_{\nu e}(p,t) & \propto  G_F^2 \int {\rm d} p_2 \, {\rm d}p_3 \,\mathbb{\Pi}_{\nu e} (p,p_2,p_3) \, {\cal F}_{\nu e}^\gtrless(\varrho(p_2,t),f_e(p_3,t),f_e(p_4,t)),\\
\mathbb{\Gamma}^\gtrless_{\nu \nu} (p,t)&  \propto G_F^2 \int {\rm d} p_2 \, {\rm d}p_3 \,\mathbb{\Pi}_{\nu \nu} (p,p_2,p_3) \, {\cal F}_{\nu \nu}^\gtrless(\varrho(p_2,t),\varrho(p_3,t),\varrho(p_4,t)).
\end{aligned}
\end{equation}
Here, $G_F$ is the Fermi constant; $p_4$ is fixed by energy-momentum conservation;  
$\mathbb{\Pi}_{\nu e}$ and $\mathbb{\Pi}_{\nu \nu}$ are scattering kernels, which are diagonal in the neutrino interaction basis; and ${\cal F}_{\nu e}^\gtrless$ and ${\cal F}_{\nu \nu}^\gtrless$ are real-time matrix products of $\varrho(p_2,t)$ with $f_e(p_3,t), f_e(p_4,t)$ and $\varrho(p_3,t), \varrho(p_4,t)$, respectively.  Because $\varrho$ can develop non-zero off-diagonal components from neutrino oscillations,  ${\cal F}^\gtrless$ and hence  $\mathbb{\Gamma}^\gtrless$ are generally not diagonal in the interaction basis.

The momentum integrals~\eqref{eq:twoD} are hard-coded in the purpose-built neutrino decoupling code {\tt FortEPianNO}~\cite{Gariazzo:2019gyi,Bennett:2020zkv}, which solves numerically the continuity equation~\eqref{eq:continuity} and the three-flavour QKEs~\eqref{eq:qke} in their entirety in precision $N_{\rm eff}^{\rm SM}$ computations.


\subsection{Damping approximation}

We would like to compute QED corrections to $\mathbb{\Gamma}^\gtrless_{\nu e}$, and estimate their impact on $N_{\rm eff}^{\rm SM}$. Ideally these corrections would take the form of corrections to the scattering kernel~$\mathbb{\Pi}_{\nu e}$, to be incorporated into a neutrino decoupling code such as {\tt FortEPianNO}.
As a first pass, however, we may work within the damping approximation, which makes the simplifying assumption that all particle species---besides that at the momentum mode $p$ represented by $\varrho(p)$---are in thermal equilibrium specified by a common temperature $T$ equal to the photon temperature.
Then, defining $\delta \varrho = \varrho(p) - \varrho_{\rm eq}(p)$, where $\varrho_{\rm eq}(p) = {\rm diag}(\fFermi(p),\fFermi(p),\fFermi(p))$ and 
$\fFermi(p) = [\exp(p/T) + 1]^{-1}$ is the Fermi-Dirac distribution, the 
collision integral~\eqref{eq:collisionintegral} can be expanded to linear order in $\delta \varrho$ to give
\begin{equation}
\label{eq:dampingapprox}
\{ {\cal I}[\varrho(p)]\}_{\alpha \beta} \simeq -\{D(p,T)\}_{\alpha \beta} \Big[\{\varrho(p) \}_{\alpha \beta} -\delta_{\alpha \beta} f_F(p)\Big],
\end{equation}
where $\delta_{\alpha \beta}$ is a Kronecker-$\delta$, and 
\begin{equation}\label{DampingCoefficients}
    \{D(p,T)\}_{\alpha \beta}\equiv \frac{1}{2} \Big[\{\mathbb{\Gamma} (p,T)\}_{\alpha \alpha} + \{\mathbb{\Gamma}(p,T) \}_{\beta \beta}    
    \Big]
\end{equation}
are the damping coefficients composed of components of the damping rate matrix
\begin{eqnarray}\label{GammaDef}
\mathbb{\Gamma} (p,T)\equiv \mathbb{\Gamma}^> (p,T)+ \mathbb{\Gamma}^< (p,T).
\end{eqnarray}
Linearisation in $\delta \varrho$ ensures that $\mathbb{\Gamma}(p,T)$ is diagonal in the neutrino interaction basis, and that $\mathbb{\Gamma}^>(p,T) = e^{E_p/T} \mathbb{\Gamma}^<(p,T)$ holds by detailed balance, where $E_p =p$ is the neutrino energy at the momentum mode~$p$ of interest.   This is the approximation under which we shall compute $\mathbb{\Gamma}$ in section~\ref{sec:calc}.  Details of the derivation of \eqref{eq:dampingapprox} can be found in, e.g., appendix~B of~\cite{Bennett:2020zkv}.

In the following, we shall devote section~\ref{sec:calc} to evaluating QED corrections to the damping rate $\{\mathbb{\Gamma}_{\nu e} (p,T)\}_{\alpha \alpha}$ due to the closed fermion loop (diagram~(d) in figure~\ref{fig:weakRates2}).  Our estimates of its impact on $N_{\rm eff}^{\rm SM}$ will be presented in section~\ref{sec:Neffestimates}.


\section{Calculation and NLO results}
\label{sec:calc}

We compute QED corrections to the interaction rates of the weak processes $\nu_\alpha e \to \nu_\alpha e$ and $\nu_\alpha \bar{\nu}_\alpha \leftrightarrow e^+ e^-$ within the framework of Fermi theory, whose effective Lagrangian can be expressed as 
\begin{align}
\label{eq:L4F}
    \mathcal{L}_{\mathrm{4F}} =   \frac{4 G_F}{\sqrt{2}} \left[\bar{\psi}_{\alpha} \gamma_\mu P_L \psi_\alpha\right]\left[g_L^\alpha J_L^\mu + g_R J_R^\mu\right].
\end{align}
Here, 
the spinors $\psi_\alpha$ represent neutrino fields of flavour $\alpha$; 
$J^\mu_{L/R} = \bar{\psi}_e \gamma^\mu P_{L/R} \psi_e$ are the left- and right-handed electron current operators, with the chiral projectors $P_{L/R}=\frac{1}{2}(1 \mp \gamma^5)$; the right-handed coupling reads $g_R = \sin^2\theta_W$, while the left-handed coupling $g^\alpha_L = -\frac{1}{2} + \sin^2\theta_W+\delta^{\alpha e}$ 
depends on the neutrino flavor $\alpha = e,\mu,\tau$.   Strictly speaking, the closed fermion loop in figure~\ref{fig:weakRates2} receives in principle contributions also from quarks.  This interaction is also well described by a Lagrangian of the form~\eqref{eq:L4F}, with the couplings $g^\alpha_L$ and $g_R$ updated for the quarks of interest.%
\footnote{The reason is that a Fierz identity has been used to bring the charged and neutral current contributions into the common form~\eqref{eq:L4F}.} 
We shall however omit the contributions from quarks in this work: at finite temperatures these contributions are Boltzmann-suppressed at the ${\cal O}(1)$~MeV temperatures of interest, while hadronic corrections in vacuum have been studied in~\cite{Hill:2019xqk}.
 As QED is a vector-like theory, we also introduce the vector-axial couplings $g^\alpha_{V,A} = \frac{1}{2}(g^\alpha_L \pm g_R)$ as an alternative notation.


\subsection{Evaluation of the closed fermion loop}

In order to compute the damping coefficients~\eqref{DampingCoefficients},
we must evaluate the flavour-diagonal entries of the neutrino damping rate matrix in the equilibrium approximation for $\varrho$, i.e., $\Gamma_\alpha(p,T) \equiv \{\mathbb\Gamma_{\nu e}(p,T)\}_{\alpha \alpha}$, which is given by the sum 
\begin{equation}
\Gamma_\alpha= \Gamma_\alpha^>(p) + \Gamma_\alpha^<(p),
\end{equation}
where $\Gamma_\alpha^\gtrless$ are the $\alpha$-flavoured neutrino production ($<$) and destruction ($>$) rates in the equilibrium limit.
In the framework of the real-time formalism of finite-temperature QED, these rates can be extracted from the Wightman self-energies $\Sigma^\gtrless$,%
\footnote{There is no term involving the plasma four-velocity $U$ because we consider the sum over helicities. See, e.g., appendix C of~\cite{Antusch:2017pkq} in the context of heavy neutrinos.} 
\begin{equation}
    \Gamma_\alpha^{\gtrless}(p) = \mp\frac{ 1}{2p^0} \left. \mathrm{Tr}\left[ 
        \slashed{p} \Sigma_\alpha^{\gtrless} \right]\right|_{p^0=E_p},
\end{equation}
where $\Sigma^\gtrless$ are identified with the  self-energies with opposite (Keldysh) contour indices:%
\footnote{The idea to evaluate fields on a closed time path was already implicit in Schwinger's seminal work~\cite{Schwinger:1960qe}. Here, we follow common convention and refer to the indices marking the forward and backward contours in the real-time formalism as Keldysh indices~\cite{Keldysh:1964ud}.}
$\Sigma^<\equiv \Sigma^{12}$ and $\Sigma^>\equiv \Sigma^{21}$.
In thermal equilibrium, detailed balance is established by the fermionic Kubo-Martin-Schwinger (KMS) relation $\Sigma^> = - e^{p^0/T} \Sigma^<$.  Then, the mode-dependent damping rate can also be written as
\begin{equation}
\label{eq:dampingGamma}
  \Gamma_\alpha (p)  =
  \frac{1}{2 p^0} \left. \mathrm{Tr}\left[\slashed{p}(\Sigma_\alpha^< - \Sigma_\alpha^>)\right]\right|_{p^0=E_p}
  = \frac{1}{2 p^0 \fFermi(p^0)} \left. \mathrm{Tr}\left[\slashed{p}\Sigma_\alpha^<\right]\right|_{p^0=E_p},
\end{equation}
where the first equality follows from the fact that $\Gamma_\alpha$ can be related to the discontinuity of the retarded neutrino self-energy evaluated at the quasiparticle pole, 
$-i{\rm disc}\Sigma^R = 2\,{\rm Im}\Sigma^R = \Sigma^> - \Sigma^<$, 
in accordance with the optical theorem at finite temperature.
While the KMS relation makes explicit use of the fact that $\Sigma^\gtrless$ are computed in thermal equilibrium, the real-time formalism used here can be generalised in a straightforward manner to non-equilibrium situations \cite{Chou:1984es}.%
\footnote{Specifically, this entails replacing the equilibrium distribution in the relation~\eqref{StatProp} between the statistical and the spectral propagators with a dynamical function, and modifying all other propagators accordingly. 
 See, e.g.~\cite{Garbrecht:2011xw,Drewes:2012qw} for a discussion.}
Where there is no confusion, we shall drop the flavour index $\alpha$ in the following.

\begin{figure}[t]
\centering
\includegraphics[width=.75\textwidth]{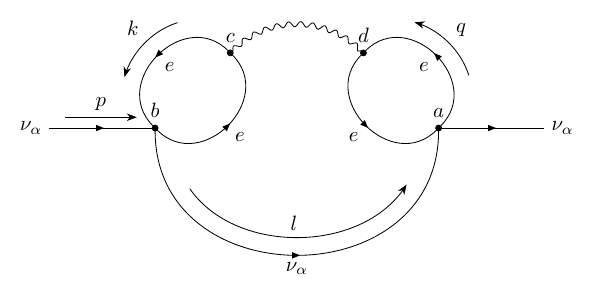}
\caption{The three-loop diagram containing the closed fermion loop. Explicitly labelled are the external ($a$ and $b$) and summed ($c$  and $d$) real-time contour indices, the momenta ($k,l,p$ and $q$) and particles ($e$ and $\nu_\alpha$).}
    \label{fig:diagram}
\end{figure}

On general grounds the dominant QED correction to  $\Gamma_\alpha$ can be expected from diagram~(d) in figure \ref{fig:weakRates2} in the regime where the photon propagator is on-shell. 
This expectation has been confirmed in \cite{Jackson:2023zkl}.
 In the real-time formalism this process is contained in the 
contribution to the Wightman self-energy shown in figure~\ref{fig:diagram}, where, in terms of finite-temperature cutting rules, diagram~(d)  can be recovered from a cut through one closed electron loop and the internal neutrino line.%
\footnote{\label{PlasmonProcessFootnote} 
Resumming the photon propagator (see discussion above equation~\eqref{eq:TthFinal} and appendix~\ref{app:1PIresummedphotonpropagator}) opens up an additional cut through the photon and the internal neutrino lines, corresponding to the so-called plasmon process $\gamma\to\nu\bar{\nu}$, even if the imaginary part of the photon self-energy is neglected. This can be seen in equation~\eqref{eq:D11resummed}, which exhibits an on-shell $\delta$-function in the limit of vanishing width~\eqref{GammaDefPhoton}. However, the plasmon process contributes at ${\cal O}(\alphaEM^2)$ (see appendix C in \cite{Braaten:1993jw}), which can be understood as a phase space suppression from $m_\gamma \sim \alphaEM T$. We therefore do not consider it in this work.}
We shall compute only this contribution and 
denote the corresponding Wightman self-energy with $\Sigma^{ba}_{\rm NLO}$.  For notational convenience, we further split $\Sigma^{ba}_{\rm NLO}$ into a sum $\Sigma^{ba}_{\rm NLO} = \sum_{c,d=1,2} \Sigma^{ba,cd}$ over the internal real-time indices $c$ and~$d$, with the partial self-energies $\Sigma^{ba,cd}$  given by
\begin{multline}
\mathrm{Tr}\left[\slashed{p}\Sigma^{ba,cd}\right] 
= (-1)^{a+b+c+d+1} 
(i e)^2 \left(\frac{i 4 G_F}{\sqrt{2}}\right)^2  \\ \times
\int_{l,q,k} \mathrm{Tr}\left[i S_e^{ad}(q)\gamma_\rho (P_L g_L^\alpha + P_R g_R) i S_e^{da}(q+p-l) \gamma^\mu\right]  i D_{\mu\nu}^{cd}(p-l) \\ \times\mathrm{Tr}\left[\slashed{p} \gamma^\rho P_L i S_{\nu_\alpha}^{ba}(l)\gamma^\sigma P_L\right]   \mathrm{Tr}\left[i S_e^{cb}(k)\gamma^\nu i S_e^{bc}(k+p-l)\gamma_\sigma (P_L g_L^\alpha + P_R g_R)\right].  
\label{eq:amp}
\end{multline}
Here, we have introduced the abbreviation $\int_p \equiv  \int \dd[4]{p}/(2\pi)^4$ for loop integrals; the traces are over the Clifford algebra; the subscripts ``$e$'' and ``$\nu_\alpha$'' on the fermionic propagators refer to the associated particle; and the definitions of the propagators can be found in appendix~\ref{sec:propagators}.

 As we would like to compute $\Sigma^{12}_{\rm NLO}$, we set the external contour indices to $a=2$ and $b=1$. Then, the contributions $\mathrm{Tr}\left[\slashed{p}\Sigma^{12,21}\right]$ and $\mathrm{Tr}\left[\slashed{p}\Sigma^{12,12}\right]$, which correspond to cutting both the photon and neutrino lines, vanish by momentum conservation (see also footnote~\ref{PlasmonProcessFootnote}). Of the remaining ``11'' and ``22'' contributions, the transformation behaviour of the thermal propagators under hermitian conjugation dictates that $\mathrm{Tr}\left[\slashed{p}\Sigma^{12,11}\right]=  \mathrm{Tr}\left[\slashed{p}\Sigma^{12,22}\right]^\ast$.  It then follows that
\begin{equation}
\label{eq:self}
\mathrm{Tr}\left[\slashed{p}\Sigma^{12}_{\rm NLO}\right] = 2 \mathrm{Re}\,\mathrm{Tr}\left[\slashed{p}\Sigma^{12,11}\right],
\end{equation}
i.e., the Wightman self-energy represented by figure~\ref{fig:diagram} can be determined entirely through the diagonal ``11'' contribution. 

It is instructive to recast the expressions for the self-energy~\eqref{eq:amp}--\eqref{eq:self} into the form of a standard Boltzmann collision integral commonly found in textbooks (e.g., \cite{Kolb:1990vq}).
To do so, we first identify the 4-momenta $p,l,q$ in figure~\ref{fig:diagram} with the momenta $p_1,p_2,p_3,p_4$ of the external neutrinos and electrons of the underlying $2 \to 2$ process represented by diagram (d) of figure~\ref{fig:weakRates2} via
\begin{align}
    p_1= p, \qquad p_2 =-l, \qquad   p_3 =q + p - l, \qquad p_4 = -q.
\end{align}
Then, writing out the propagators $iS_e^{12/21}$ and $iS_\nu^{12/21}$ explicitly, the self-energy~\eqref{eq:amp}--\eqref{eq:self} can be brought into the form 
\begin{equation}
\mathrm{Tr}\left[\slashed{p}\Sigma^{12}_{\mathrm{NLO}}\right] = -(2\pi)^3\int_{q,l} \delta(p_2^2) \delta(p_3^2 - m_e^2) \delta(p_4^2 - m_e^2)\mathcal{F}(p_2^0,p_3^0,p_4^0) 
    \mathcal{T}_{\rm NLO},
    \label{eq:amp_simple}
\end{equation}
where the population factor
\begin{equation}
    \mathcal{F}(p_2^0,p_3^0,p_4^0) = \left[\fFermi(|p_2^0| )-\Theta (p_2^0)\right]  \left[\fFermi(|p_3^0| )-\Theta (-p_3^0)\right] \left[\fFermi(|p_4^0| )-\Theta (-p_4^0)\right]
\end{equation}
 contains the equilibrium phase space distributions of the three integrated external particles and is analogous to $\cal{F}^\gtrless$ in equation~\eqref{eq:twoD}.  The quantity ${\cal T}_{\rm NLO}$ plays the role of QED corrections to the LO squared matrix element,%
\footnote{For completeness, $\mathcal{T}_{\mathrm{LO}} = 2^7 G_F^2 \left[(g_L^\alpha)^2 \left(p_1\cdot p_4\right)^2 +g_R^2   \left(p_2\cdot p_4\right)^2 +g_L^\alpha g_R  m_e^2
   \left(p_1\cdot p_2\right)\right]$.  If we were to replace ${\cal T}_{\rm NLO}$ with ${\cal T}_{\rm LO}$ in equation~\eqref{eq:amp_simple}, we would recover the leading-order equilibrium neutrino production rates.}
and can be written in terms of the one-loop photon self-energy $\Pi_{ab}^{\mu\nu}$  as
 \begin{multline}
    \mathcal{T}_{\rm NLO} =  -16 G_F^2 g_V^\alpha  \mathrm{Re}\left\{
   \mathrm{Tr}\left[(\slashed{p}_3+m_e)\gamma^\mu \
   (m_e-\slashed{p}_4)\gamma^\rho\left(g_L^\alpha  P_L + g_R P_R\right)\right] \right. \\ \left. \times  \mathrm{Tr}\left[\slashed{p}_1  \gamma_\rho P_L\slashed{p}_2\gamma_\sigma P_L\right] D^{11}_{\mu\nu}(P) \Pi_{11}^{\nu\sigma}(P)\right\},
   \label{eq:TNLO}
\end{multline}
where we have introduced the photon momentum $P = p_1 + p_2$.

Since ${\cal T_{\rm NLO}}$ has the interpretation of a squared matrix element, we can split it into a vacuum and a thermal part according to temperature dependence, 
\begin{equation}
{\cal T}_{\rm NLO}= {\cal T}_{\rm vac} + {\cal T}_{\rm th}.
\end{equation}
The vacuum correction ${\cal T}_{\rm vac}$ has no intrinsic temperature dependence in the sense that it makes no explicit reference to the temperature or to any phase space distribution.  It is simply the correction to the weak matrix elements
arising from the interference of the closed fermion loop diagram~(d) with the LO graph in standard $T=0$ quantum field theory, and can be expressed in terms of the vacuum photon self-energy as
 \begin{align}
       \mathcal{T}_{\rm vac} =  -
         2^{10} G_F^2\, \alphaEM \, [(g^\alpha_V)^2/(4 \pi)] \left[ m^2_e (p_1\cdot p_2) 
            + 2 (p_1\cdot p_4)^2 \right]\mathrm{Re}\, \Pi_{2}(P^2)
        \label{eq:Tvac}
 \end{align}
where  $\alpha_{\rm em}$  is the electromagnetic fine-structure constant, and the form factor~$\Pi_2$ is defined in appendix~\ref{app:1PIresummedphotonpropagator}; see equation \eqref{eq:Pi_2}. The simplicity of the expression follows from the fact that the integration domain is symmetric under the interchange $p_3\leftrightarrow p_4$.  This symmetry, along with momentum conservation, also ensures the absence of all antisymmetric terms containing Levi-Civita symbols arising from traces of $\gamma^5$ with four or more $\gamma$-matrices.  Vacuum QED corrections to the neutrino-electron interaction matrix element were previously computed in reference~\cite{Tomalak:2019ibg}.

The thermal correction ${\cal T}_{\rm th}$, on the other hand, depends explicitly on equilibrium phase space distributions, $f_{F}$ or $f_B$, of the internal particles (``$F$'' for Fermi-Dirac; ``$B$'' for Bose-Einstein),
and can be thought of as a temperature-dependent correction to the squared matrix element. This $T$-dependence originates in the thermal part of the ``11''
propagator, which is proportional to $\delta(p^2-m^2) f_{F/B}(|p^0|)$ (see appendix~\ref{sec:propagators}) and, where it is applied, effectively puts an internal line of the closed fermion loop diagram (d) in figure~\ref{fig:weakRates2} on-shell.  Purely from counting, there are altogether seven possible ways to put one, two, or all three internal lines of diagram~(d) on-shell. Not all combinations contribute to ${\cal T}_{\rm NLO}$: Terms proportional to $f_B$ correspond to putting the photon line on-shell,  
which are forbidden for diagram~(d) for kinematic reasons~\cite{Braaten:ApJ,Braaten:1993jw}. 
Similarly, putting both internal electrons on-shell leads to a purely imaginary contribution that is irrelevant to the real part of the self-energy we wish to compute.  The only surviving two combinations correspond to putting either internal electron line on-shell, and are proportional to $\fFermi(|k^0|)$ and $\fFermi(|k^0+P^0|)$ respectively.

Observe that ${\cal T}_{\rm NLO}$ contains
a $t$-channel contribution from  elastic $\nu_\alpha e$ scattering (i.e., $p_2^0<0$) that is logarithmically divergent for soft photon momenta.  This divergence comes from the fact that the finite-temperature photon self-energy
scales not as $P^2$ like in vacuum, but as $T^2$ in the hard-thermal loop limit which do not compensate anymore for the $1/P^2$ behaviour of the photon propagator. In addition, soft photons are Bose-enhanced.   To remedy the problem,
 we replace the tree-level photon propagator $D^{cd}_{\mu\nu}$ in equation~\eqref{eq:TNLO} with the fully-resummed photon propagator $\bar{D}^{ab}_{\mu\nu}$.%
\footnote{In principle, it is also necessary to replace the tree-level photon propagator in equation~\eqref{eq:Tvac} with the fully resummed one. However, since the vacuum correction is not dominated by small photon momenta, the choice of photon propagator has in this case a negligible impact on the  numerical outcome.}
Furthermore, because both $\bar{D}^{ab}_{\mu\nu}$ and $\Pi_{11}^{\nu\sigma}$ split into a longitudinal (``$L$'') and a transverse (``$T$'') part, the same decomposition applies also to ${\cal T}_{\rm th}$, i.e., $\mathcal{T}_{\rm th} =  \mathcal{T}^{L}_{\rm th} +  \mathcal{T}^{T}_{\rm th}$, where $\mathcal{T}^{L,T}_{\rm th}$ can  be brought into the form 
\begin{multline}
      \mathcal{T}^{L/T}_{\rm th} = -2^8 G_F^2 (g_V^\alpha)^2 \mathrm{Re}\, \bar{D}^{L/T}_{R}(P)  \mathrm{Re}\,\Pi^{L/T}_{R,T\neq 0}(P)\left[   2 (p_1\cdot p_4) P^{1,4}_{L/T}
       \right. 
       \\  \left.  
       +  (p_1\cdot p_2)(a_{L/T} (p_1\cdot p_2) + P^{2,2}_{L,T} + P^{4,4}_{L/T})  \right].
       \label{eq:TthFinal}
\end{multline}
Here, $\bar{D}_R$ denotes the retarded resummed photon propagator; $\Pi_{R, T\neq 0}$
is the retarded thermal photon self-energy comprising the $\fFermi(|k^0|)$ and $\fFermi(|k^0+P^0|)$ terms described above; 
we have employed the shorthand notation $P^{i,j}_{L/T} = P^{\mu\nu}_{L/T} p_{i,\mu} p_{j,\nu}$; and $a_{L,T} = (3 \mp 1)/4$ further differentiates between the longitudinal and the transverse contribution. 

Note that the imaginary part of the photon propagator,  $\mathrm{Im} \, \bar{D}^{L,T}_{11}$, does not appear in equation~\eqref{eq:TthFinal} because it is formally of higher-order in $\alphaEM$ and we are only interested in the $\order{\alphaEM}$ corrections.
We also only use the resummed photon propagator in the IR divergent $t$-channel contribution; where the divergence is absent, i.e., in the $s$-channel, we set $\bar{D}_R\to D_R$, where $D_R$ is the un-resummed counterpart of $\bar{D}_R$.
 The final expressions for $\mathrm{Re}\,\Pi_{11,T\neq 0}^{L/T}$ and  $\mathrm{Re}\, \bar{D}^{L/T}_{11}$ are given in equations~\eqref{eq:PiL11HTL}, \eqref{eq:PiT11HTL} and~\eqref{eq:D11resummed}, which can be easily mapped to $\mathrm{Re}\,\Pi_{R,T\neq 0}^{L/T}$ and $\mathrm{Re}\, \bar{D}_R$ via $\mathrm{Re}\, \bar{D}^{L/T}_{R}= \mathrm{Re}\, \bar{D}^{L/T}_{11}$ and $\mathrm{Re}\,\Pi^{L/T}_{R} = \mathrm{Re}\,\Pi^{L/T}_{11}$.  Details of their derivation, along with the relevant thermal integrals, can be found in appendices~\ref{app:Kintegrals} and \ref{app:1PIresummedphotonpropagator}.


\subsection{Numerical results for NLO neutrino interaction rate}

We evaluate the self-energy correction~\eqref{eq:amp_simple} and hence the correction to the neutrino damping rate~\eqref{eq:dampingGamma} by numerically integrating over the two remaining momenta $l$ and $q$ in~\eqref{eq:amp_simple} using the parameterisation shown in appendix~\ref{app:param}.  We adopt the experimentally-determined values given by the Particle Data Group~\cite{ParticleDataGroup:2022pth} for the following input parameters:
\begin{itemize}
    \item Fermi's constant: $G_F = \SI{1.1663788\pm0.0000006}{\mega\electronvolt^{-2}}$,
    \item Electron mass: $m_e = \SI{0.51099895000\pm 0.00000000015}{\mega\electronvolt}$,
    \item Electromagnetic fine-structure constant: $\alphaEM^{-1}(0) = \SI{137.035999180(10)}{} $, and
    \item Weinberg angle: $\sin^2 \theta_W(0)_{\MSbar}=  \SI{0.23863\pm 0.00005}{}$.
\end{itemize}
The renormalisation scale $\mu_R$ appearing in the photon self-energy of the vacuum contribution is identified with the electron mass, $\mu_R = m_e$.

\begin{figure}[t]
    \centering
    \includegraphics[width=0.65\textwidth]{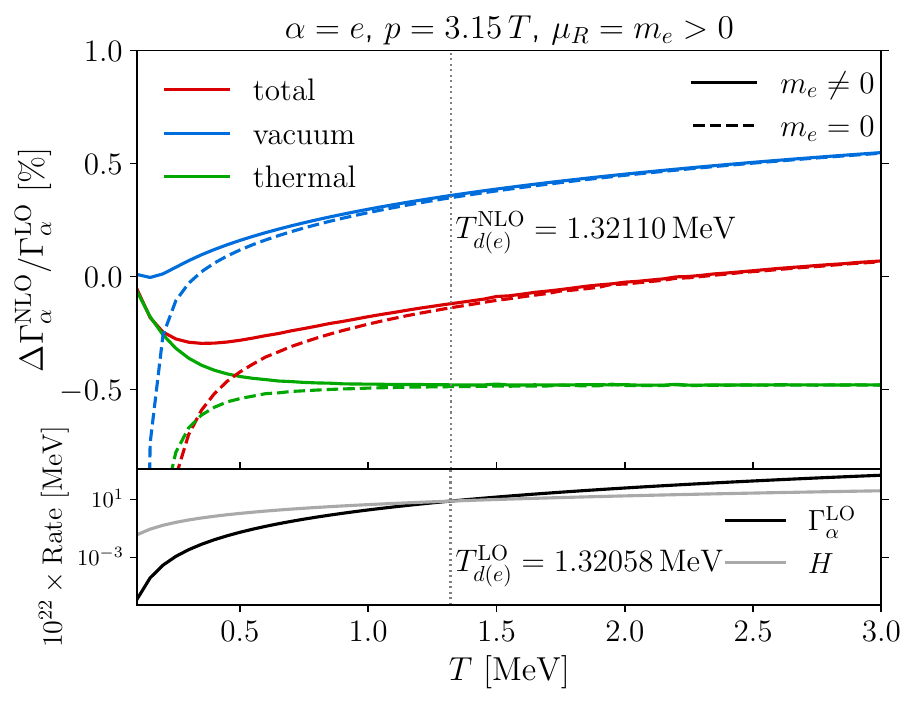}
   \hspace*{-5mm} \includegraphics[width=0.67\textwidth]{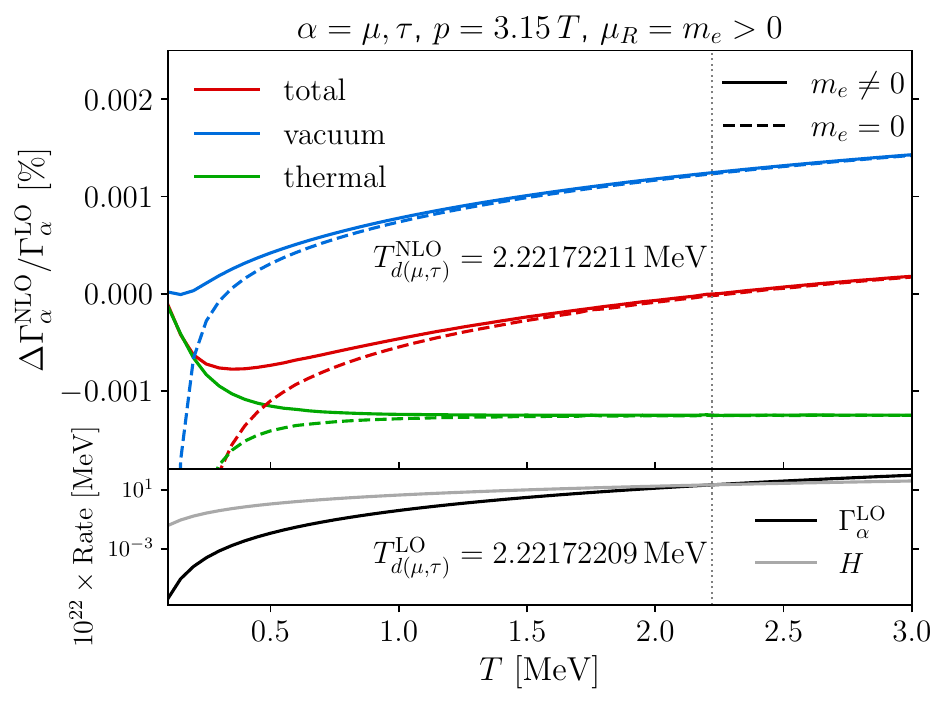}
    \caption{{\it Top}: NLO contributions to the $\nu_e$ interaction rate from the closed fermion loop with a finite electron mass (solid) or $m_e=0$ (dashed) at different  temperatures for the mean neutrino momentum $p=3.15 T$. For comparison we normalise all curves to the LO rate (without QED corrections), which we always evaluate with $m_e\neq0$ to ensure a common normalisation.  
   The total rate correction (red) is further split into the vacuum (blue) and the thermal contribution (green). 
   At $T\gg m_e$, the green curve flattens out as the thermal correction contains no scale besides the temperature in this region, while the vacuum correction retains a mild dependence on the renormalisation scale~$\mu_R$.  
   The lower panel shows the LO neutrino interaction rate compared to the Hubble rate, and $T_d$ indicates the decoupling temperature, defined via $\Gamma_\alpha(T_{d(e)}) = H(T_{d(e)})$. 
    {\it Bottom}: Same as top, but for $\nu_{\mu,\tau}$. 
    \label{fig:GnuNLO}
    }
\end{figure}

Figure~\ref{fig:GnuNLO} shows the closed fermion loop corrections to the damping rates $\Gamma_e(p)$ and $\Gamma_{\mu,\tau}(p)$ at the mean momentum $p=3.15 T$. Relative to their respective LO rates, we find the corrections at temperatures $T \sim 1 \to 3$~MeV to fall in the range $-0.2 \to +0.1\%$ and $-0.0005 \to +0.0002\%$, respectively, for $\nu_e$ and $\nu_{\mu,\tau}$.
We further note that:
\begin{enumerate}
\item  At $T \sim \SI{2}{\mega\electronvolt}$, the vacuum and the thermal contributions to the correction are roughly equal in magnitude (green vs blue lines in figure \ref{fig:GnuNLO}), in contrast to the findings of~\cite{Cielo:2023bqp}, where finite-temperature corrections were determined to be subdominant. We note however that
a direct comparison is not possible because a 
different set of diagrams was  investigated in~\cite{Cielo:2023bqp}---(a), (b), and (c) in figure~\ref{fig:weakRates2}---as opposed to the type~(d) corrections considered in this work.

\item 
Reference~\cite{Cielo:2023bqp} also found no significant flavour dependence in the  rate corrections: their $\Gamma_e$ and $\Gamma_{\mu,\tau}$ corrections differ by less than $1\%$ in the temperature regime around neutrino decoupling.  Our corrections are on the other hand strongly flavour-dependent---by more than two orders of magnitude---and trace their origin to the fact that electron neutrinos experience charge current interactions while the muon- and tau-flavoured neutrinos interact only via the neutral current with the $e^\pm$ thermal bath.  This difference renders the corresponding vector couplings, $g_V^e \sim 0.49$ and $g_V^{\mu,\tau} \sim -0.012$, very roughly two orders of magnitude apart from one another.  This strong flavour dependence  in the rate corrections was also seen in reference~\cite{Jackson:2023zkl}. 

\item   
We have computed the NLO contributions to the interaction rates in two different approximations: (i)~retaining the full dependence on the electron mass (solid lines in figure \ref{fig:GnuNLO}), and (ii)~in the limit $m_e \rightarrow 0$ (dashed lines). The massless calculation aims to quantify the effect of the $m_e=0$ approximation used in reference~\cite{Jackson:2023zkl}, along with the Hard Thermal Loop (HTL) approximation of the photon propagator. 

We observe that the error from neglecting $m_e$ is relatively minor for $T \gtrsim 3m_e$, but becomes sizeable at low temperatures. In particular, in the limit $T\rightarrow 0$ the ratio $\Gamma_\alpha^{\rm NLO}/\Gamma_\alpha^{\rm LO}$ vanishes for $m_e > 0$, but diverges for $m_e = 0$ because of Boltzmann suppression of the LO rates. Precision computations of $N_{\rm eff}^{\rm SM}$ track the evolution of neutrinos down to temperatures much below $m_e$~\cite{Froustey:2020mcq,Bennett:2020zkv}.  Thus, although it is commonly understood that (electron) neutrino decoupling occurs at relativistic temperatures $T \sim (2 \to 3) \times m_e$, what we assume for $m_e$ in the NLO rate computations may yet have some impact on $N_{\rm eff}$ (see section~\ref{sec:boltzmannNeff}). 

\end{enumerate}

\begin{figure}[t]
    \centering
    \includegraphics[width=0.65\textwidth]{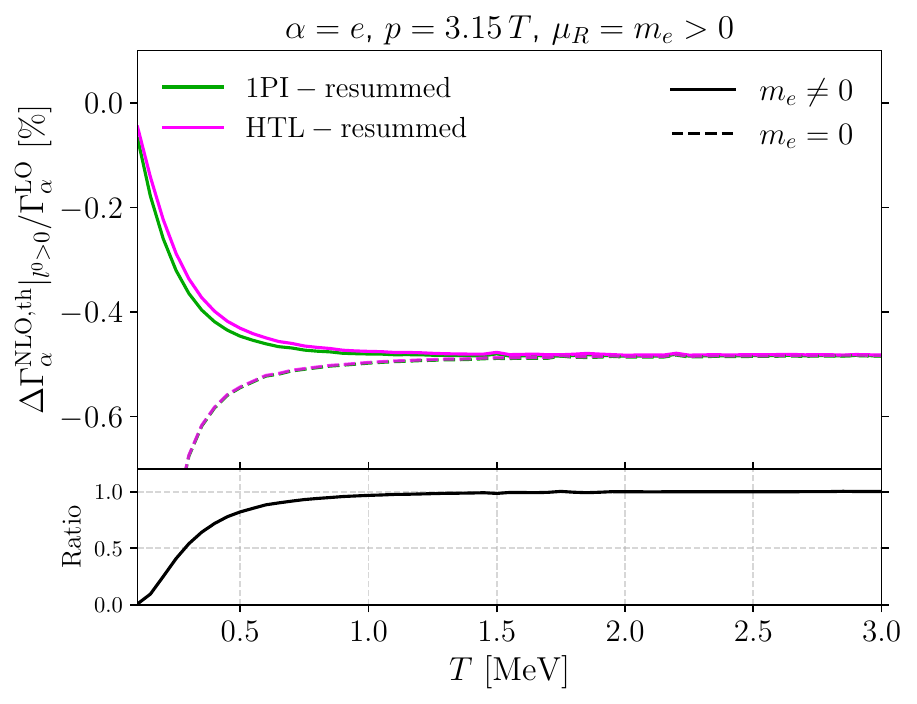}
  \hspace*{-6mm}  \includegraphics[width=0.68\textwidth]{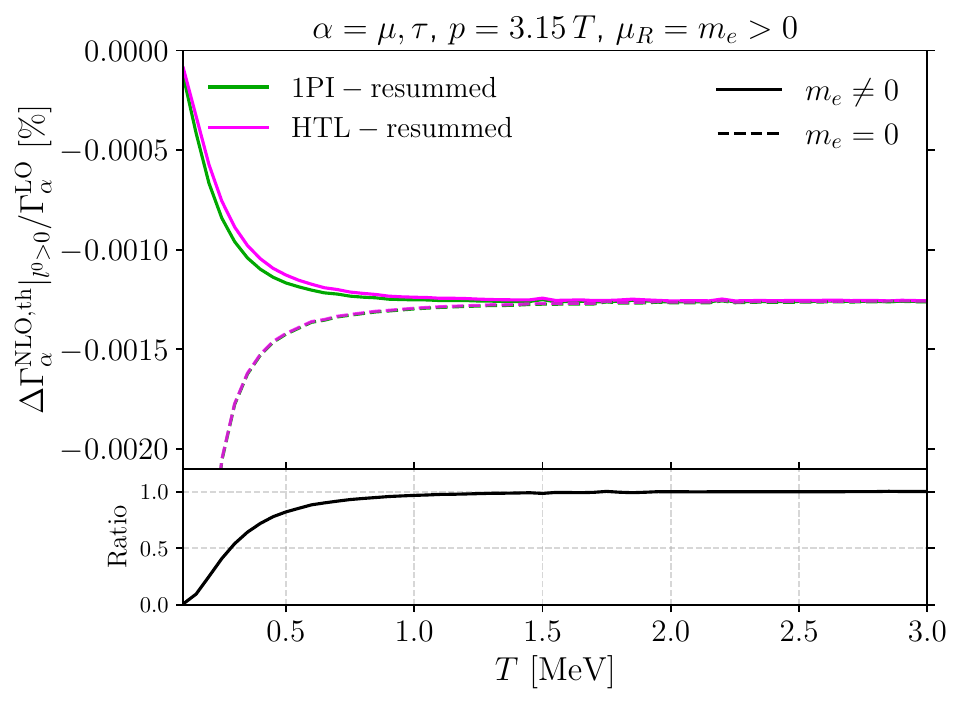}
    \caption{\emph{Top:} The $t$-channel contribution ($l^0>0$) to the NLO electron neutrino interaction rate 
    at the mean momentum $p=3.15T$
    in different approximations, namely, using 1PI-resummed (green) and HTL (magenta) photon propagators, in each case with $m_e\neq 0$ (solid lines) or $m_e = 0$ (dashed lines) in both electron loops of the self-energy diagram of figure~\ref{fig:diagram}. As in figure \ref{fig:GnuNLO}, we normalise all curves to the LO rate, which we always evaluate with $m_e\neq0$.
    The lower panel shows the ratio of the 1PI result for $m_e\neq 0$ to the HTL result for $m_e = 0$ as a function of the temperature. \emph{Bottom}: Same as top, but for $\alpha\neq e$.}
    \label{fig:HTLcomp}
\end{figure}

 Figure~\ref{fig:HTLcomp} focuses on the $t$-channel contribution to the interaction rate, where the enhancement near the photon mass-shell occurs.  We show four sets of results, based on the resummed photon propagator detailed in appendix~\ref{app:1PIresummedphotonpropagator}: 
 (i)~the complete one-loop result including a finite $m_e$ everywhere,
 (ii)~the complete one-loop result in the limit $m_e \rightarrow 0$, 
 (iii)~using the HTL photon propagator (which does not depend on the electron mass), but with $m_e$ everywhere else,%
 \footnote{Even though it is physically inconsistent to keep $m_e$ in the electron propagators and not in the photon one, (iii) is nonetheless helpful as it isolates the dependence of the rate corrections to the 1PI-resummed vs the HTL propagator.} 
 and (iv)~using the HTL photon propagator and setting $m_e=0$ everywhere. 
 As expected, we observe that (ii) and (iv)  match to a very good approximation. Indeed, since the scattering rates are dominated by the kinematic region around the $t$-channel singularity where photons are soft, the 1PI-resummed propagator is well-approximated by the HTL one when, in addition, we set $m_e = 0$.  On the other hand, visible differences can be discerned between (i) and (iii) at $T \lesssim 1$~MeV, which can be explained by the fact that the HTL approximation only holds for $T\gg m_e$. In the lower panel of figure~\ref{fig:HTLcomp}, we highlight the impact of $m_e$ by displaying the ratio of (i) to (iv).


\subsection{NLO decoupling temperatures}

The lower panels of the two plots in figure~\ref{fig:GnuNLO} show the LO 
electron neutrino and muon/tau neutrino interaction rates juxtaposed with the Hubble expansion rate.  The latter is given by $H^2(T) = (\rho_{\rm QED} + \rho_\nu)/(3 M^2_{\rm Pl})$, where $M_{\rm Pl} = \SI{2.43536e21}{\mega\electronvolt}$ is the reduced the Planck mass, $\rho_{\nu}  = 3 g_\nu (7/8)(\pi^2/30) T^4$  is the energy density in three families of neutrinos with $g_\nu=2$, and
\begin{align}
     \rho_{\rm QED} = g_\gamma \frac{\pi^2 }{30} T^4 +  \frac{g_e}{2 \pi^2} \int_0^\infty  \dd{p} p^2 E_e \fFermi(E_e)
\end{align}
is the energy density in the QED plasma assuming an ideal gas configuration, with $g_\gamma=2$, $g_e =4$, and $E_e = \sqrt{p^2 + m_e^2}$. By solving the equation $\Gamma_\alpha^{\rm LO}(T_{d(\alpha)}) = H(T_{d(\alpha)})$ for the flavour-dependent decoupling temperatures $T_{d(\alpha)}$, we determine the LO decoupling temperatures to be
\begin{equation}
\begin{aligned}
  &  T^{\mathrm{LO}}_{d(e)} \simeq 
  \SI{1.32058}{\mega\electronvolt},  \\
  &  T^{\mathrm{LO}}_{d(\mu,\tau)}  \simeq 
\SI{2.22172209}{\mega\electronvolt}.
    \label{eq:tdLO}
\end{aligned}
\end{equation}
 The first number differs from the earlier estimate of $T^{\mathrm{LO}}_{d(e)} \simeq 1.3453$~MeV from reference~\cite{Bennett:2019ewm}, which might be traced to different choices of input parameter values. Incorporating QED corrections to the damping rates, the decoupling temperatures shift to 
 \begin{equation}
 \begin{aligned}
    &T^{\mathrm{NLO}}_{d(e)} \simeq 
     \SI{1.32110}{\mega\electronvolt},\\
    &T^{\mathrm{NLO}}_{d(\mu,\tau)} \simeq 
    \SI{2.22172211}{\mega\electronvolt}
    \label{eq:TdNLO}
\end{aligned}
\end{equation} 
corresponding to an increase of $\sim \SI{0.04}{\percent}$ for $\nu_e$ and of  $\sim \SI{8e-07}{\percent}$ for $\nu_{\mu,\tau}$.  Around the muon neutrino decoupling temperature, the vacuum and thermal contributions to the NLO rates approximately cancel, explaining the smallness of the correction to $T^{\mathrm{NLO}}_{d(\mu,\tau)}$.

Given that \cite{Jackson:2023zkl} computed the NLO weak rates assuming $m_e = 0$, it is also of interest to study how such an assumption modifies the decoupling temperature shifts.
Using our  computations of the $m_e=0$ rate corrections (but with $m_e \neq 0$ LO rates), we find the corresponding NLO decoupling temperatures to be
\begin{equation} 
 \begin{aligned}
    &T^{\mathrm{NLO},m_e=0}_{d(e)} \simeq  \SI{1.32118}{\mega\electronvolt}, \\
    &T^{\mathrm{NLO},m_e=0}_{d(\mu,\tau)} \simeq  \SI{2.22172223}{\mega\electronvolt} ,
    \label{eq:TdNLOme=0_new}
\end{aligned}
\end{equation}
i.e., a $\SI{0.05}{\percent}$ and  $\SI{6e-06}{\percent}$ shift for $\nu_e$ and $\nu_{\mu,\tau}$, respectively, relative to their corresponding LO decoupling temperature~\eqref{eq:tdLO}.

\section{NLO effects on \texorpdfstring{$N_{\rm eff}^{\rm SM}$}{Neff}}
\label{sec:Neffestimates}

Having  computed the closed fermion loop correction to the damping rate $\Gamma_\alpha$, we are now in a position to estimate its effect on $N_{\rm eff}^{\rm SM}$.  Within the damping approximation, two avenues are available to us:
\begin{itemize}
\item Given $\Gamma_\alpha$ at a representative momentum, we have already estimated in equation~\eqref{eq:TdNLO} the corresponding correction to the neutrino decoupling temperature $T_d$, defined via $\Gamma_{\alpha}(T_d) = H(T_d)$. Then, under the assumption of instantaneous neutrino decoupling, an estimate of the change to $N_{\rm eff}^{\rm SM}$, $\delta N_{\rm eff} \equiv N_{\rm eff}^{\rm NLO}-N_{\rm eff}^{\rm LO}$, can be obtained via entropy conservation arguments.

\item We may also compute $\delta N_{\rm eff}$ by solving directly the continuity equation~\eqref{eq:continuity} and the QKEs~\eqref{eq:qke} in the damping approximation~\eqref{eq:dampingapprox}  and neglecting neutrino oscillations.

\end{itemize}
We consider both approaches in the following.  The corresponding estimates for $\delta N_{\rm eff}$ are summarised in table~\ref{tab:dNeff}.

\begin{table}[t]
\centering
\begin{tabular}{l c c}
\toprule
Method & $\delta N_{\rm eff}/N^{\rm LO}_{\rm eff}$ & $\delta N_{\rm eff}^{m_e = 0}/N^{\rm LO}_{\rm eff}$ \\
\midrule
Entropy conservation (common decoupling) & $\SI{-1.1e-05}{}$ & $\SI{-1.2e-05}{}$ \\ \addlinespace[1mm]
Entropy conservation (flavour-dependent decoupling) & $\SI{-5.4e-06}{}$ & $\SI{-6.1e-06}{}$ \\  \addlinespace[1mm]
Boltzmann damping approximation (mean momentum)  &  $\SI{-7.8e-06}{}$ &  $\SI{-1.6e-05}{}$ \\ 
 \addlinespace[1mm]
Boltzmann damping approximation (full momentum)  &  $\SI{-7.9e-06}{}$ & $\SI{-2.6e-05}{}$  \\ 
\bottomrule
\end{tabular}
\caption{Estimates of the relative correction to $N^{\rm SM}_{\rm eff}$ due to NLO weak rate corrections, with and without the electron mass, using different methods.}
\label{tab:dNeff}
\end{table}


\subsection{Entropy conservation}
\label{sec:entropy}

The entropy conservation argument posits that entropies in a comoving volume in two decoupled sectors are separately conserved, i.e., 
\begin{equation}
\begin{aligned}
\label{eq:entropy}
s(a_1) a_1^3 & = s(a_2) a_2^3, \\
s_{\nu_\alpha}(a_1) a_1^3 & = s_{\nu_\alpha}(a_2) a_2^3,
\end{aligned}
\end{equation}
where $s_{\nu_\alpha}$ and $s \equiv s_\gamma + s_e + \sum_{\beta\neq \alpha} s_{\nu_\beta}+\cdots$ denote, respectively the entropy density of a decoupled neutrino species $\nu_\alpha$ and of the QED plasma plus any neutrino species $\nu_\beta$ that may still be in equilibrium with it, and the scale factors~$a_{1,2}$ represent two different times after decoupling.  We take $a_1$ to correspond to the time immediately after $\nu_\alpha$ decouples instantaneously from the QED plasma, i.e., $a_1 = a_{d}^+$, where the sector temperatures satisfy $T(a_1) = T_{\nu_\alpha}(a_1) \equiv T_d$, while $a_2 > a_1$ is some later time to be specified below.

The assumption of instantaneous decoupling allows the neutrinos to maintain to an excellent approximation an ideal-gas equilibrium phase space distribution at all times, so that $s_{\nu_\alpha} (a) \propto T_{\nu_\alpha}^3(a)$.  It then follows simply from~\eqref{eq:entropy} that $T_{\nu_\alpha}(a_2) = (a_1/a_2) T_{\nu_\alpha}(a_1)=(a_1/a_2) T_d$.    In the case of the QED+$\nu_\beta$ plasma, we parameterise its entropy density as
\begin{equation}
 s (a) = \frac{2 \pi^2}{45} \tilde{g}_s (a) T^3(a),
\end{equation}
where the entropy degree of freedom parameter $\tilde{g}_s$ is given in the ideal gas limit by
\begin{equation}
\tilde{g}_s (a)\equiv g_\gamma  + \frac{45}{4 \pi^4}\frac{g_e}{T^4(a)} \int_0^\infty {\rm d}p\, p^2 \left(E_e + \frac{p^2}{3 E_e} \right) \fFermi(E_e,T(a)) + \frac{7}{8} \sum_{\beta \neq \alpha} g_{\nu_\beta},
\end{equation}
with $g_\gamma=2$, $g_e=4$, and $g_{\nu_\beta} = 2$. For our current purpose of estimating $\delta N_{\rm eff}$ due to NLO contributions to the weak rates, it suffices to use the ideal-gas $\tilde{g}_s$.  We do note however that QED entropy density at $T_d\sim {\cal O}(1)$~MeV is subject in principle to a sizeable finite-temperature correction to the QED equation of state, which needs to be included in precision $N_{\rm eff}^{\rm SM}$ calculations.  See, e.g., reference~\cite{Bennett:2019ewm} for details.  

Then, combining the above, we find an estimate of the neutrino-to-photon temperature ratio of
\begin{equation}
\label{eq:temp}
\frac{T_{\nu_\alpha}(a_2)}{T(a_2)} = \left[\frac{\tilde{g}_s(a_2)}{\tilde{g}_s(a_d^+)} \right]^{1/3}
\end{equation}
at the later time $a_2$. 
We use this temperature ratio in the following to provide two estimates of $\delta N_{\rm eff}$ due to the rate corrections.


\paragraph{Common decoupling temperature}  In the first estimate, we assume all neutrino flavours to decouple effectively  at the same time, an assumption that may to some extent be justified by the observed large mixing in the neutrino sector.  We choose $a_1=a_d^+$ to correspond to the time immediately after $\nu_e$ decoupling, and set $a_2$ at a time significantly after $e^\pm$ annihilation, where $T(a_2) \ll m_e$.  The latter leads immediately to an entropy degree of freedom of $\tilde{g}_s(a_2) = g_\gamma =2$, while for the former we have
\begin{equation}
\tilde{g}_s (a_d^+)= 2 + \frac{45}{\pi^4 T_{d(e)}^4} \int_0^\infty {\rm d}p\, p^2 \left(E_e + \frac{p^2}{3 E_e} \right) \fFermi(E_e,T_{d(e)}).
\end{equation}
  Then, using the temperature ratio~\eqref{eq:temp} and the ideal-gas relations $\rho_\gamma \propto g_\gamma T^4$ and $\rho_{\nu_\alpha} \propto (7/8) g_{\nu_\alpha} T_{\nu_\alpha}^4$, we find
\begin{equation}
\label{eq:rhoratio}
\left.\frac{\rho_\nu}{\rho_\gamma}\right|_{T/m_e \to 0} = \sum_\alpha \frac{\rho_{\nu_\alpha}(a_2)}{\rho_\gamma(a_2)}= 3 \times \frac{7}{8}  \left[\frac{2}{\tilde{g}_s (a_d^+)} \right]^{4/3},
\end{equation}
or, equivalently in terms of $N_{\rm eff}$,
\begin{equation}
N_{\rm eff} = 3 \times  \left[\frac{11}{4}\frac{2}{\tilde{g}_s(a_d^+)} \right]^{4/3}, 
\end{equation}
where the factor $4/11$ corresponds to $2/\tilde{g}_{s}(T_d)$ evaluated in the limit $T_d \gg m_e$.  

Using the LO and NLO electron neutrino decoupling temperatures in~\eqref{eq:tdLO} and \eqref{eq:TdNLO} respectively, we find a fractional shift of $\delta N_{\rm eff}/N_{\rm eff}^{\rm LO} \simeq  \SI{-1.1e-05}{}$ due to the rate corrections.
Had we instead used the $m_e=0$ NLO decoupling temperatures~\eqref{eq:TdNLOme=0_new}, the correspond shift in $N_{\rm eff}$ would have been  $\delta N_{\rm eff}^{m_e=0}/N_{\rm eff}^{\rm LO}\simeq  \SI{-1.2e-05}{}$.  Thus, while setting $m_e=0$ leads to a $\sim 10\%$ change in the estimate of $\delta N_{\rm eff}$, its ultimate impact on $N_{\rm eff}^{\rm SM}$ appears to be small, at least within the entropy conservation picture.  
These results are summarised in table~\ref{tab:dNeff}.

\paragraph{Flavour-dependent decoupling}  In the absence of neutrino oscillations, $\nu_e$ and $\nu_{\mu,\tau}$ decouple at different temperatures, $T_{d (e)}$ and $T_{d(\mu,\tau)}>T_{d(e)}$. Then, to estimate $\delta N_{\rm eff}$ requires that we consider entropy conservation across four epochs:
the time immediately after $\nu_{\mu,\tau}$ decoupling $a_1=a_{d(\mu,\tau)}^+$; immediately {\it before} $\nu_e$ decoupling $a_2=a_{d(e)}^-$; immediately {\it after} $\nu_e$ decoupling  $a_3=a_{d(e)}^+$; and $a_4$ is a time significantly after $e^\pm$ annihilation. The corresponding entropy degrees of freedom are
\begin{equation}
\begin{aligned}
\tilde{g}_s (a_{d(\mu,\tau)}^+) &= 2 + \frac{45}{\pi^4 T_{d(\mu,\tau)}^4} \int_0^\infty {\rm d}p\, p^2 \left(E_e + \frac{p^2}{3 E_e} \right) \fFermi(E_e,T_{d(\mu,\tau)}) + \frac{7}{8} g_{\nu_e},\\
\tilde{g}_s (a_{d(e)}^-) &= 2 + \frac{45}{\pi^4 T_{d(e)}^4} \int_0^\infty {\rm d}p\, p^2 \left(E_e + \frac{p^2}{3 E_e} \right) \fFermi(E_e,T_{d(e)}) + \frac{7}{8} g_{\nu_e},\\
\tilde{g}_s (a_{d(e)}^+) &= 2 + \frac{45}{\pi^4 T_{d(e)}^4} \int_0^\infty {\rm d}p\, p^2 \left(E_e + \frac{p^2}{3 E_e} \right) \fFermi(E_e,T_{d(e)}),\\
\end{aligned}
\end{equation}
and $\tilde{g}_s(a_4) =2$. An estimate of the $\nu_e$-to-photon energy density ratio at $a_4$ follows straightforwardly from the temperature ratio~\eqref{eq:temp} and ideal-gas temperature-energy relations: 
\begin{equation}
\label{eq:rhoratioe}
\frac{\rho_{\nu_e}(a_4)}{\rho_\gamma(a_4)}=\frac{7}{8}\left[\frac{\tilde{g}_s(a_4)}{\tilde{g}_s(a_{d(e)}^+)} \right]^{4/3} =\frac{7}{8}\left[\frac{2}{\tilde{g}_s(a_{d(e)}^+)} \right]^{4/3}.
\end{equation}
For $\rho_{\nu_{\mu,\tau}}(a_4)/\rho_\gamma(a_4)$, we note that $\nu_e$ decoupling at $a_{d(e)}$ introduces a discontinuity in $\tilde{g}_s$, thereby leading to a more complicated 
energy density ratio at $a_4$,
\begin{equation}
\label{eq:rhoratiomutau}
\frac{\rho_{\nu_{\mu,\tau}}(a_4)}{\rho_\gamma(a_4)} = \frac{7}{8}\left[\frac{2}{\tilde{g}_s(a_{d(e)}^+)} \frac{\tilde{g}_s(a_{d(e)}^-)}{\tilde{g}_s(a_{d(\mu,\tau)}^+)} \right]^{4/3}.
\end{equation}
Then, combining equations~\eqref{eq:rhoratioe} and~\eqref{eq:rhoratiomutau} to form $\rho_\nu = \sum_\alpha \rho_{\nu_\alpha}$, we find
\begin{equation}
\label{eq:neffflavour}
N_{\rm eff} = \left(\frac{11}{4}\frac{2}{\tilde{g}_s (a_{d(e)}^+)} \right)^{4/3} \left[1+ 2 \times \left(\frac{\tilde{g}_s(a_{d(e)}^-)}{\tilde{g}_s(a_{d(\mu,\tau)}^+)}\right)^{4/3}
\right] \, 
\end{equation}
for our estimate.

Evaluating~\eqref{eq:neffflavour} for the LO and NLO decoupling temperatures~\eqref{eq:tdLO} and~\eqref{eq:TdNLO}, we find a fractional shift in $N_{\rm eff}$ of  $\delta N_{\rm eff}/N_{\rm eff}^{\rm LO} \simeq \SI{-5.4e-06}{}$ if we use the $m_e\neq 0$ corrections, or $\delta N^{m_e=0}_{\rm eff}/N_{\rm eff}^{\rm LO} \simeq \SI{-6.1e-06}{}$ using the $m_e=0$ corrections.  Compared with the common-decoupling estimates displayed in table~\ref{tab:dNeff}, we note that the flavour-dependent decoupling estimates of $\delta N_{\rm eff}/N_{\rm eff}^{\rm LO}$ are generally about a factor of two smaller, in both the $m_e=0$ and $m_e \neq 0$ cases.  This difference is to be expected, as the QED corrections to the $\nu_{\mu,\tau}$ interaction rates are negligible compared with the corrections to the $\nu_e$ rates.  We emphasise, however, both estimates are very crude approximations: the true shift in $\delta N_{\rm eff}$ will probably fall somewhere in-between.


\subsection{Solving the neutrino Boltzmann equations and the continuity equation}
\label{sec:boltzmannNeff}

Following reference~\cite{Mangano:2001iu}, we introduce the comoving quantities $x\equiv m_e\, a$,  $y\equiv a\, p$, and $z \equiv T\, a$, and rewrite  
 the continuity equation~\eqref{eq:continuity} as a differential equation for the quantity~$z$ (corresponding to the photon-to-neutrino temperature ratio):
\begin{align}
\dv{z}{x} &= \frac{\frac{x}{z}J(x/z)-\frac{1}{2 z^3}  \dv{\bar{\rho}_\nu}{x} +G_1(x/z)}
{\frac{x^2}{z^2}J(x/z)+Y(x/z)+\frac{2\pi^2}{15}+G_2(x/z)}\, ,
\label{eq:dz/dx}
\end{align}
where $J(x/z)$ and $Y(x/z)$ describe the ideal-gas behaviour of the QED plasma, while the $G_{1,2}(\tau)$ account for deviations of the QED plasma from an idea gas.  Explicit forms for these expression to ${\cal O}(e^2)$ can be found in appendix~\ref{sec:continuitysupplement}.

Equation~\eqref{eq:dz/dx} also requires as input the total time derivative of the comoving neutrino density $\bar{\rho}_\nu \equiv {\rho}_\nu a^4$, which can generally be constructed from the neutrino density matrices via
\begin{equation}
\label{eq:drhobar/dx}
  \frac{{\rm d} \bar{\rho}_\nu}{{\rm d} x} = \frac{1}{\pi^2}   \int \mathrm{d} y~ y^3 \sum_\alpha \dv{\{\varrho(x,y)\}_{\alpha\alpha}}{x},
\end{equation}
where ${\rm d} {\{\varrho(x,y)\}_{\alpha\alpha}}/{\rm d} x$ corresponds in general to the QKEs~\eqref{eq:qke}.  Neglecting neutrino oscillations, the density matrices $\varrho$ are diagonal in the flavour basis, in which case the QKEs~\eqref{eq:qke} simplify to a set of classical Boltzmann equations.  Then, together with the damping approximation~\eqref{eq:dampingapprox} and in terms of the new variables, we find
\begin{align}
\label{eq:rhoyeqn}
\dv{\{\varrho(x,y)\}_{\alpha\alpha}}{x} & \simeq  -\frac{\Gamma_\alpha(x,y)}{xH(x)}\Big[\{\varrho(x,y)\}_{\alpha\alpha}-\fFermi((y/z)T)\Big].
\end{align}
Equation~\eqref{eq:rhoyeqn} can be solved together with the continuity equation~\eqref{eq:dz/dx} for a range of momenta~$y$ covering the bulk of the neutrino population (a typical range would be $y \in [ 0.01, 30]$). We call this the ``full-momentum'' approach. 
Alternatively, we can simplify the equation~\eqref{eq:rhoyeqn}  further by adopting the {\it ansatz}
\begin{equation}
   \{ \varrho(y)\}_{\alpha\alpha} = \{\varrho(\langle y \rangle) \}_{\alpha\alpha} \frac{\fFermi(yT)}{\fFermi(\langle y \rangle T)},
\end{equation}
and $\Gamma_\alpha(y) = \Gamma_\alpha (\langle y \rangle)$.  Identifying $\langle y \rangle$ with the mean momentum mode  $y_0 = 3.15\, z(T_0)$, where  $T_0$ is the photon temperature at initialisation (which is equal to the neutrino temperature), we can rewrite equation~\eqref{eq:drhobar/dx} in this approximation as
\begin{equation}
  \frac{{\rm d} \bar{\rho}_\nu}{{\rm d} x} =- \frac{7\pi^2}{120 xH(x)} \left[\sum_\alpha \Gamma_\alpha(\langle y \rangle) \left( \frac{\{\varrho(x,\langle y \rangle)\}_{\alpha\alpha}}{\fFermi(\langle y \rangle T)}-z^3(x)\right)\right].
\end{equation}
We call this alternative the ``mean-momentum'' approach.

Irrespective of whether we use the full-momentum or the mean-momentum approach, the final $N_{\rm eff}$ can be estimated from the solutions to $\rho_\alpha$ and $z$ at $x \to \infty$ using the definition of $N^{\rm SM}_{\rm eff}$ in equation~\eqref{NeffFermief}, reproduced here in terms of the rescaled variables: 
\begin{equation}
N_{\rm eff} = \left. \frac{8}{7} \left(\frac{11}{4}\right)^{4/3} \frac{30}{2 \pi^2}\, \left[\frac{z(T_0)}{z(x)}\right]^{4} \bar{\rho}_\nu (x)\right|_{x \to \infty}.
\end{equation}
Table~\ref{tab:dNeff} shows our estimates of $\delta N_{\rm eff}/N_{\rm eff}^{\rm LO}$ due to QED corrections to the neutrino interaction rates using both the full-momentum and the mean-momentum approaches, with or without the electron mass in the correction.

Evidently from table~\ref{tab:dNeff}, both the full-momentum and the mean-momentum estimates for $m_e \neq 0$, $\delta N_{\rm eff}/N_{\rm eff}^{\rm LO} \simeq (-7.8 \to -7.9) \times 10^{-6}$, are broadly consistent with their counterpart obtained in section~\ref{sec:entropy} from entropy arguments: in fact they sit between the common-decoupling and flavour-dependent decoupling estimates from entropy considerations.  In contrast, the estimates assuming $m_e = 0$ differ by $\sim 50\%$ between the full-momentum ($\delta N_{\rm eff}^{m_e=0}/N_{\rm eff}^{\rm LO} \simeq -2.6 \times 10^{-5}$) and the mean-momentum ($\delta N_{\rm eff}^{m_e=0}/N_{\rm eff}^{\rm LO} \simeq -1.6 \times 10^{-5}$) approaches, and are furthermore $\sim 30\%$ to a factor of four larger than those from entropy arguments.  This result is consistent with expectations.  As we have seen previously in figures~\ref{fig:GnuNLO} and~\ref{fig:HTLcomp}, the rate correction under the $m_e=0$ assumption diverges as $T \to 0$ relative the LO rate, whereas its $m_e\neq 0$ counterpart tends to zero.  Since neutrino decoupling in the early universe is not instantaneous but extends into the $e^\pm$-annihilation epoch at $T \sim m_e$, any estimate of $\delta N_{\rm eff}$ that accounts for non-instantaneous decoupling will to some extent be sensitive to exactly what we assume for $m_e$ in the $T \lesssim 3 m_e$ region.  Indeed, the low-$T$ effect of $m_e$ on $\delta N_{\rm eff}/N_{\rm eff}^{\rm LO}$ is not captured by entropy conservation arguments, which are based on one point estimate (the decoupling temperature) at $T> m_e$, where the $m_e\neq0$ and $m_e=0$ rate corrections differ by less than 10\%.  In contrast, the full-momentum approach, which is sensitive to the largest range of temperatures, yields the strongest dependence of $\delta N_{\rm eff}/N_{\rm eff}^{\rm LO}$ on what we assume for the electron mass.
Thus, while the overall effect of the rate corrections on $N_{\rm eff}^{\rm SM}$ is small, we conclude that neglecting $m_e$ in their computations is strictly not a good approximation in high-precision calculations of $N_{\rm eff}^{\rm SM}$.


\section{Conclusions}
\label{sec:conclusions}

In this work we have computed the QED corrections to the neutrino-electron interaction rate in the vicinity of neutrino decoupling and evaluated its impact on
the standard-model value of the effective number of neutrinos~$N_{\rm eff}^{\rm SM}$. We have focused on diagram~(d) in figure~\ref{fig:weakRates2}, because of the expectation of a $t$-channel enhancement and hence dominance over the other three diagrams.  Similar corrections 
have also been recently investigated in Cielo~{\it et al.}~\cite{Cielo:2023bqp} and Jackson and Laine~\cite{Jackson:2023zkl}, which the former analysis~\cite{Cielo:2023bqp} found to lead to a significant shift in $N_{\rm eff}^{\rm SM}$: from the benchmark value of $3.044$~\cite{Froustey:2020mcq,Bennett:2020zkv} to $3.043$.

Contrary~\cite{Cielo:2023bqp}, our first-principles calculations show that QED corrections to the neutrino interaction rates are modest.  In the temperature range $T \sim 1 \to 3$~MeV, the corrections to electron neutrino interaction rate fall in the range $-0.2 \to +0.1\%$ relative to the LO rate, while for the muon and tau neutrinos the effect is even more minute: $-0.0005 \to +0.0002\%$.  These results are consistent with those reported by Jackson and Laine~\cite{Jackson:2023zkl}, despite differences between the formalisms (imaginary vs real-time) and some approximations (zero vs finite $m_e$) used.
The more than two orders of magnitude difference between the relative corrections for the $\nu_e$ and the $\nu_{\mu,\tau}$ rates also confirms the strong flavour dependence found in~\cite{Jackson:2023zkl},  which was not observed in Cielo~{\it et al.}~\cite{Cielo:2023bqp}.

Using our QED-corrected neutrino interaction rates, we proceed to estimate the corresponding shift in $N_{\rm eff}^{\rm SM}$ under a variety of approximations and methods: via entropy conservation arguments which assume instantaneous decoupling, and by solving the Boltzmann equation in the damping approximation.  Depending on the exact method/approximation used, we find the relative change in $N_{\rm eff}^{\rm SM}$ to fall in the range  $\delta N_{\rm eff}/N_{\rm eff}^{\rm LO} \simeq (-0.5 \to - 1.1) \times 10^{-5}$.  That is, relative to the current SM benchmark of $N_{\rm eff}^{\rm SM}= 3.0440 \pm 0.0002$~\cite{Froustey:2020mcq,Bennett:2020zkv}, QED corrections to the neutrino interaction rates can at best shift the number in the negative direction in the fifth decimal place, and are hence completely within the quoted uncertainties.  Thus, while we can confirm the sign of $\delta N_{\rm eff}$ computed by Cielo {\it et al.}~\cite{Cielo:2023bqp}, even our most ``optimistic'' estimate is a factor of 30 smaller than their claimed correction.

 It is also interesting to observe that setting $m_e=0$ in the rate corrections can have an ${\cal O}(1)$ impact on $\delta N_{\rm eff}/N_{\rm eff}^{\rm LO}$, even though the rate corrections themselves at $T \sim 1\to3$~MeV differ by less than  $10\%$.  This is because corrections assuming $m_e=0$ deviate from their $m_e \neq 0$ counterparts at $T \lesssim 3 m_e$, and diverge relative to LO rates just as the $m_e \neq 0$ corrections vanish in the $T \to 0$ limit.  Since neutrino decoupling in the early universe is not instantaneous, these $T \lesssim 3 m_e$ effects will imprint on $N_{\rm eff}^{\rm SM}$ despite the common understanding that neutrino decoupling occurs at $T \sim 1$~MeV.
Thus, while the overall contribution from QED weak rate corrections to $N_{\rm eff}^{\rm SM}$ is ultimately small, we argue that neglecting $m_e$ their computations is strictly speaking not a good approximation in high-precision $N_{\rm eff}^{\rm SM}$ calculations. 

In conclusion, our results strongly suggest that the SM benchmark value
$N_{\rm eff}^{\rm SM}= 3.0440 \pm 0.0002$~\cite{Froustey:2020mcq,Bennett:2020zkv}
is correct within the quoted uncertainties.  Naturally, a full numerical solution of the QKEs by a dedicated neutrino decoupling code such as {\tt FortEPiaNO}~\cite{Gariazzo:2019gyi}---with all NLO contributions from diagrams (a) to (d) incorporated in the collision integral---would be highly desirable.  However, short of a new and previously unaccounted-for effect, we believe it is unlikely that a more detailed investigation of NLO effects on the neutrino interaction rate will yield a deviation from the existing SM benchmark $N_{\rm eff}^{\rm SM}$ that is large enough to be of any relevance for cosmological observations in the foreseeable future.

\acknowledgments  
MaD and Y$^3$W would like to thank Isabel Oldengott and Jamie McDonald for discussions. 
 YG acknowledges the support of the French Community of Belgium through the FRIA grant No.~1.E.063.22F.  The research of MK and LPW was supported by the Deutsche Forschungsgemeinschaft (DFG, German Research Foundation) through the Research Training Group 2149 ``Strong and Weak Interactions - from Hadrons to Dark matter''.
Y$^3$W is supported in part by the Australian Research Council’s Future Fellowship (project FT180100031) and Discovery Project (project DP240103130).
Computational resources have been provided by the University of Münster through the HPC cluster PALMA II funded by the DFG (INST 211/667-1), and the supercomputing facilities of the Université Catholique de Louvain (CISM/UCL) and the Consortium des Équipements de Calcul Intensif en Fédération Wallonie Bruxelles (CÉCI) funded by the Fond de la Recherche Scientifique de Belgique (F.R.S.-FNRS) under convention 2.5020.11 and by the Walloon Region.

\appendix

\section{Bosonic and fermionic propagators at finite temperatures}
\label{sec:propagators}

For vanishing chemical potentials the free thermal propagators for fermions are given by
\begin{equation}
\begin{aligned}
  i S_\psi^{11}
  &= \frac{i(\slashed{p}+m_\psi)}{p^2-m_\psi^2+i\epsilon} - 2\pi (\slashed{p}+m_\psi)\delta(p^2-m_\psi^2)\fFermi(|p^0|) = \gamma^0 (i S_\psi^{22})^\dag \gamma^0,\\
  i S_\psi^{12} &= - 2\pi (\slashed{p}+m_\psi)\delta(p^2-m_\psi^2)\left[\fFermi(|p^0|) - \Theta(-p^0)\right], \\
  i S_\psi^{21} &= - 2\pi (\slashed{p}+m_\psi)\delta(p^2-m_\psi^2)\left[\fFermi(|p^0|) - \Theta(p^0)\right],
\end{aligned}
\end{equation}
while the ones for bosons are
\begin{equation}
\begin{aligned}
     i \Delta^{11}  &=  \frac{i}{p^2-m^2+i\epsilon} +  2\pi \delta(p^2 - m^2) f_B(|p^0|) = (i \Delta^{22} )^\ast,\\
     i \Delta^{12} &= 2\pi \delta(p^2-m^2)\left[f_B(|p^0|) + \Theta(-p^0)\right],  \\
     i \Delta^{21} &= 2\pi \delta(p^2-m^2)\left[f_B(|p^0|) + \Theta(p^0)\right],  
\end{aligned}
\end{equation}
where $f_B(E)=1/(e^{E/T}-1)$ is the Bose-Einstein distribution.

For the computation of the resummed photon propagator, it is useful to write down the advanced and retarded bosonic propagators,
\begin{align}
  i\Delta^{R/A} = \frac{i}{p^2-m^2\pm i \mathrm{sgn}(p^0) \epsilon},
\end{align}
as well as the statistical propagator
\begin{equation}\label{StatProp}
    i\Delta^+ \equiv \frac{i}{2}(\Delta^{12}+\Delta^{21}) = \pi  \left[1 + 2 f_B(|p^0|)\right]\delta(p^2 - m^2) = \frac{1}{2} \left[1 + 2 f_B(p^0)\right]\Delta^-,
\end{equation}
where the relation to the spectral function $\Delta^- = i(\Delta^{21} - \Delta^{12})$ is another way of stating the KMS relation.
In terms of the scalar propagators, the tree-level photon propagator is given by
\begin{equation}
    i D^{ab}_{\mu \nu}(p) = i d_{\mu\nu} \Delta^{ab}(p)|_{m=0},
    \label{eq:treelevelphotonpropagator}
\end{equation}
where we have defined the polarisation tensor,
\begin{equation}
  d^{\mu\nu}(p) = -g^{\mu\nu} + (1-\xi)\frac{p^\mu p^\nu}{p^2} ,
\end{equation}
for a general $R_\xi$-gauge.

\section{Thermal integrals}
\label{app:Kintegrals}

Throughout the calculations we encounter thermal integrals of the following forms
\begin{equation}
\begin{aligned}
  &K^\mu(P) = 2 e^2 \int \frac{\dd[4]{k}}{(2\pi)^3} k^\mu \frac{\delta(k^2 - m_e^2) \fFermi(|k^0|)}{(P+k)^2 - m_e^2 + i \epsilon},\\
  &K^{\mu\nu}(P) = 2 e^2 \int \frac{\dd[4]{k}}{(2\pi)^3} k^\mu k^\nu \frac{\delta(k^2 - m_e^2) \fFermi(|k^0|)}{(P+k)^2 - m_e^2 + i \epsilon}.
  \label{eq:Kmunu}
\end{aligned}
\end{equation}
To evaluate them, we choose $\mathbf{P}$ as the reference direction with the corresponding unit vector~$\hat{\mathbf{P}}$, and parameterise the integration variable via
\begin{equation}
  \mathbf{k} = \ak ( \cos\vartheta \hat{\mathbf{P}} + \cos\varphi \sin\vartheta \mathbf{e}_1 +  \sin\varphi \sin\vartheta \mathbf{e}_2 ),
  \label{eq:param_k}
\end{equation}
where, together, $\hat{\mathbf{P}}$ and the two unit vectors $\mathbf{e}_1$ and $\mathbf{e}_2$ form a complete orthogonal basis.

The contributions from $\mathbf{e}_1$ and $\mathbf{e}_2$ to the spatial components of $K^\mu$ vanish after the azimuthal integration, giving 
\begin{equation}
  \mathbf{K} = \frac{ \mathbf{P}}{\aP^3}\frac{\alphaEM }{4\pi} \int_{m_e}^\infty \dd{\omega}  \left[ P^2 \ell_1(\omega,P) + 2 \omega P_0  \ell_2(\omega,P) - 8  \ak \aP \right] \fFermi(\omega),
\end{equation}
where  $\omega = |k^0| = \sqrt{\ak^2 + m_e^2}$, and 
\begin{equation}
\begin{aligned}
  & \ell_1(\omega,P)=  \ln \left|\frac{(P^2 + 2 \ak \aP)^2 - 4 P_0^2 \omega^2}{(P^2 - 2 \ak \aP)^2 - 4 P_0^2 \omega^2}\right| , \\ 
  & \ell_2(\omega,P)= \ln \left| \frac{P^4  - 4 ( P_0 \omega +  \aP \ak)^2 }{P^4  - 4 ( P_0 \omega -  \aP \ak)^2} \right|   
  \label{eq:logfunc}
\end{aligned}
\end{equation}
arise from integration over $\cos\vartheta$. Correspondingly,
\begin{equation}
  K^0 = \frac{\alphaEM}{2 \pi \aP} \int_{m_e}^\infty  \dd{\omega} \omega \, \ell_2(\omega,P) \fFermi(\omega) 
\end{equation}
is the remaining, temporal component of $K^\mu$.

For $K^{\mu\nu}$, after application of the completeness relation $\delta^{ij} = \hat{P}^i \hat{P}^j + \sum_{k=1,2} e_k^i e_k^j$, the spatial components become
\begin{equation}
  K^{ij} = \hat{P}^i \hat{P}^j G_1 + (\delta^{i j} - \hat{P}^i \hat{P}^j) (I_2 - I_1/2 ),
\end{equation}
 where we have introduced the two auxiliary functions
 \begin{equation}
\begin{aligned}
  &I_1 = \frac{\alphaEM}{8\pi \aP^3} \int_{m_e}^\infty \dd{\omega}  \left[ (P^4 + 4 P_0^2 \omega^2) \ell_1(\omega,P)  + 4 P^2 P_0 \, \omega \,  \ell_2(\omega,P) - 8 \aP |\mathbf{k}| P^2 \right] \fFermi(\omega), \\
  &I_2 = \frac{\alphaEM}{4\pi \aP } \int_{m_e}^\infty \dd{\omega} \ak^2    \ell_1(\omega,P) \fFermi(\omega).
\end{aligned}
\end{equation}
The remaining components with at least one index being zero are
\begin{equation}
\begin{aligned}
  &K^{00}(P) = \frac{\alphaEM}{2 \pi \aP} \int_{m_e}^\infty  \dd{\omega} \omega^2 \, \ell_1(\omega,P) \fFermi(\omega), \\
  &K^{0i}=  \frac{\alphaEM \hat{P}^i}{4\pi \aP^2} \int_{m_e}^\infty \dd{\omega} \omega  \left[ 2 P_0  \omega   \ell_1(\omega,P) + P^2  \ell_2(\omega,P)\right] \fFermi(\omega).
\end{aligned}
\end{equation}
In all cases, the remaining single integral over $\omega$ can, for our purposes, be evaluated numerically efficiently.


\section{1PI-resummed photon propagator}
\label{app:1PIresummedphotonpropagator}

A crucial ingredient in our calculation is the resummed photon propagator used in our NLO thermal matrix element~\eqref{eq:TthFinal}. In this appendix, we provide an exact derivation of the photon self-energy at the one-loop level within the real-time formalism, assuming electrons in thermal equilibrium with zero chemical potentials. From there, we extract the resummed Feynman propagator. While exact expressions for this resummed photon propagator have been long known in the simplified scenarios wherein $m_e = 0$~\cite{Peshier:1998dy} or in the HTL approximation (see, e.g.,~\cite{Carrington:1997sq,Bellac:2011kqa}), here we remove these assumptions and derive a propagator valid for all values of the electron mass $m_e$ and photon 4-momentum $P$. We  note that, while the reference~\cite{Braaten:1993jw} also studied the photon propagator without HTL approximations (and accounting for non-zero electron chemical potentials), no explicit formulae were provided.

\subsection{1-loop photon self-energy}

\begin{figure}[t]
    \centering
    \includegraphics[width=.6\textwidth]{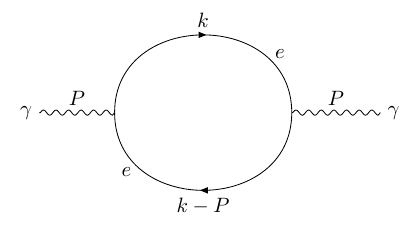}
    \caption{One-loop contribution to the photon self-energy.}
    \label{fig:photonselfenergy}
\end{figure}

At the one-loop level, the thermal photon self-energy, displayed in figure~\ref{fig:photonselfenergy}, reads%
\footnote{We only compute here the electron contribution to the photon self-energy, which is by far the dominant contribution at the relevant temperatures for $N_{\mathrm{eff}}$. It is nonetheless straightforward to extend our results to higher temperatures where different charged particles can also contribute.}
\begin{equation}
  \Pi^{\mu\nu}_{ab}(P) = (-1)^{a+b} i e^2 \int \frac{\dd[4]{k}}{(2\pi)^4} \mathrm{Tr}\left[
    i S^{ab}_e(k) \gamma^{\mu} i S_e^{b a}(k - P)
    \gamma^{\nu} \right]
    \label{app:full_Pi}
\end{equation}
for general Keldysh indices $a,b \in \{1,2\}$. It can always be split into a longitudinal and a transverse part,
\begin{equation}
  \Pi_{ab}^{\mu\nu} = \Pi^L_{ab} P_{L}^{\mu\nu} + \Pi^T_{ab} P_{T}^{\mu\nu},
  \label{eq:decomp_Pi}
\end{equation}
with the corresponding projectors~\cite{Landsman:1986uw,Rebhan:2001iw}
\begin{equation}
\begin{aligned}
  &P_T^{00}=P_T^{0i}=P_T^{i0}=0,~~  P_{T}^{i j} = -\delta^{i j } +  \hat{P}^i \hat{P}^j , \\
  &P_{L}^{\mu\nu} = g^{\mu\nu} - \frac{P^\mu P^\nu}{P^2} - P_{T}^{\mu\nu}. 
\end{aligned}
 \label{eq:projectors}
\end{equation}
Writing the self-energy in this form makes also manifest that $\Pi^{\mu\nu}_{ab}$ fulfills the Ward-Takahashi identity $P_\mu \Pi^{\mu\nu}_{ab} = 0$.
The longitudinal projector can alternatively be expressed through the heat bath four-velocity $U^\mu = \delta^\mu_0$ (in its rest frame) as $P_{L}^{\mu\nu} = \tilde{U}^\mu \tilde{U}^\nu/{\tilde{U}^2}$ with $\tilde{U}^\mu= P^2 U^\mu - (U\cdot P) P^\mu$, so that the transverse projector becomes $P_{T}^{\mu\nu} = g^{\mu\nu} - \frac{P^\mu P^\nu}{P^2} - P_{L}^{\mu\nu}$. 

In the following, we only highlight the derivation of $\Pi_{11}$ and $\Pi_{12}$, as the remaining self-energies can be related to the first two via the bosonic KMS 
relation
\begin{align}
    \Pi_{21}(P) = e^{P_0/T}\Pi_{12}(P)
    \label{eq:KMSrelation}
\end{align}
and
\begin{align}
    \Pi_{22}(P) = -\Pi_{11}(P)^{*}
\end{align}
for a general four-vector $P$.

Starting with $\Pi^{\mu\nu}_{11}$ ($\Pi^{\mu\nu}_{22}$), its diagonal component 
splits into the vacuum self-energy 
\begin{align}
  \Pi^{\mu\nu}_2 = i e^2 \int_k  \frac{i}{k^2 - m_e^2+i \epsilon} \frac{i}{(k+P)^2 - m_e^2+i \epsilon} \mathrm{Tr}\left[(\slashed{k} + m_e)  \gamma^\mu  (\slashed{k} + \slashed{P} + m_e)  \gamma^\nu \right]
\end{align}
and the finite-temperature part $\Pi_{11,T\neq 0}^{\mu \nu}$ ($\Pi_{22,T\neq 0}^{\mu \nu}$). The renormalised vacuum part $\Pi^{\mu\nu}_2(P) =  (\alphaEM/\pi) (P^2 g^{\mu\nu} - P^\nu P^\mu )  \Pi_2(P^2)$ is textbook material and, in the \MSbar-scheme, reads
\begin{equation}
\begin{aligned}
  \Pi_2(P^2) &= \frac{2}{3} \ln(\frac{\mu}{m_e}) + \frac{5}{9} + \frac{\tau }{3} + \frac{1}{3}\left(1 +  \frac{\tau }{2}\right) \sqrt{1 - \tau} \ln(\frac{\sqrt{1 - \tau} - 1}{\sqrt{1 - \tau} + 1}) \\
  &\longrightarrow\frac{1}{3} \ln(\frac{\mu^2}{-P^2}) + \frac{5}{9} \ \ \text{for} \ m_e\to 0,
  \label{eq:Pi_2}
\end{aligned}
\end{equation}
with $\tau = 4 m_e^2/P^2$ (see, e.g., appendix A of~\cite{Vanderhaeghen:2000ws} for the derivation). In practice, we note that the vacuum contribution to the photon self-energy is numerically irrelevant when resumming the photon propagator, as it vanishes in the soft photon exchange limit for which $P^2 = 0$. We can therefore safely neglect it in the resummed propagator.

In terms of the integrals $K^\mu$ and $K^{\mu\nu}$ computed in appendix~\ref{app:Kintegrals}, the corresponding real parts of the two diagonal components read
\begin{align}
  \mathrm{Re}\,\Pi_{11,T\neq 0}^{\mu \nu} &= -\mathrm{Re}\,\Pi_{22,T\neq 0}^{\mu \nu} = 4 (-g^{\mu\nu} K\cdot P + 2 K^{\mu\nu} + K^\mu P^\nu + K^\nu P^\mu) .
\end{align}
After collecting all the terms according to the projectors defined in equation \eqref{eq:projectors}, we obtain for the longitudinal part
\begin{align}
     \mathrm{Re}\,\Pi_{11,T\neq 0}^{L} &= \frac{\alphaEM P^2}{\pi \aP^3} \int_{m_e}^\infty \dd{\omega} \left[8 \ak \aP - \ell_1(\omega,P) (P^2 + 4 \omega^2) - 4 \ell_2(\omega,P) P_0 \omega\right] \fFermi(\omega) \label{eq:PiL11full} \\
     &\HTLeq -3 m_\gamma^2 \left(1 - \frac{P_0^2}{\aP^2}\right) \left[1 - \frac{P_0}{2 \aP} \ln \left| \frac{ P_0 +  \aP \ }{ P_0  -  \aP} \right|\right] \longrightarrow -3 m_\gamma^2 \ \ \text{for} \ P_0=0, \label{eq:PiL11HTL}  
\end{align}
and for the transverse one 
\begin{align}
     \mathrm{Re}\,\Pi_{11,T\neq 0}^{T} &= \frac{\alphaEM}{\pi \aP^3} \int_{m_e}^\infty \dd{\omega} \left[2 \ell_2(\omega,P) P_0 P^2 \omega -4 \ak \aP (P_0^2 + \aP^2) \right. \nonumber \\ &\hspace{2cm}\left.  + \ell_1(\omega,P) ( \aP^2 P^2+2 P_0^2 \omega^2-2 \ak^2 \aP^2+\frac{1}{2} P^4)\right] \fFermi(\omega) \label{eq:PiT11full} \\
     &\HTLeq - \frac{3}{2}m_\gamma^2 \frac{P_0^2}{\aP^2} \left[1 - \left(1 - \frac{\aP^2}{P_0^2}\right) \frac{P_0}{2\aP} \ln \left| \frac{ P_0 +  \aP \ }{ P_0  -  \aP} \right|\right]  \longrightarrow 0,  \ \ \text{for} \ P_0=0 \label{eq:PiT11HTL}
\end{align}
with the thermal photon mass $m_\gamma=e T/3$. Here, the symbol $\HTLeq$ refers to the result obtained in the HTL approximation \cite{Braaten:1989mz}, defined through the assumption that the quantities in the integral can be separated into hard $\order{\pi T}$ and soft scales $\order{eT}$, and assuming that hard momenta $k$ dominate the loop integral. 
The external momentum $P$ on the other hand is defined to be soft and therefore also negligible compared to the loop momentum. More precisely, to go from the fully momentum- and mass-dependent expression~\eqref{eq:PiL11full} (or \eqref{eq:PiT11full}) to the HTL result~\eqref{eq:PiL11HTL} (or \eqref{eq:PiT11HTL}), we have expanded in $P_0/|\textbf{k}|$ at the first step and then in $P_0/|\textbf{P}|$ in the second step, using also
\begin{align}
  \ell_1(\omega,P) \HTLeq -2\frac{\aP}{\ak}, \ \ \ \ell_2(\omega,P) \HTLeq 2 \ln \left| \frac{ P_0 +  \aP \ }{ P_0  -  \aP} \right|
\end{align}
in the same expansion scheme.  We further assume that $m_e\ll T$, such that the electron can be treated as effectively massless. This leaves us with an integral over $\omega$ that can be performed analytically using the relation $\int_0^\infty \dd{k} k\, \fFermi(k)= \pi^2 T^2/12$, thereby yielding the HTL result~\eqref{eq:PiL11HTL} (or \eqref{eq:PiT11HTL}).

The (purely imaginary) off-diagonal component $\Pi^{\mu\nu}_{12}$, on the other hand, reads
\begin{equation}
\begin{aligned}
  \Pi^{\mu\nu}_{12} =  i(i e)^2 \int & \frac{\dd[4]{k}}{(2\pi)^4} 16 \pi ^2 \delta \left(k^2-m_e^2\right) (\fFermi(| k^0| )-\Theta (k^0)) \delta \left((k+P)^2-m_e^2\right) \\ & \times (\fFermi(| k^0+P^0| )-\Theta (-k^0-P^0)) \,
 T^{\mu \nu},
\end{aligned}
\end{equation}
 where the minus sign from the fermionic loop and the one from the fact that we have a type 1 and a type 2 vertex cancel out, and  
\begin{equation}
 T^{\mu\nu} = -g^{\mu \nu } k\cdot P +2 k^{\mu } k^{\nu } +k^{\nu } P^{\mu
  } + k^{\mu }P^{\nu }.
\end{equation}
For the evaluation of the integrals, we make use again of the decomposition of $\mathbf{k}$ shown in equation \eqref{eq:param_k}. The additional $\delta$-function here (compared to the diagonal case) fixes then the polar angle $\theta$ to the value
\begin{equation}
  \cos\theta^\ast_\pm = \frac{P^2 \pm 2 \omega P_0}{2 \ak \aP} .
\end{equation}
With that, $T^{\mu\nu}$ separates into a transverse and a longitudinal part according to
\begin{align}
   T_\pm^{\mu\nu} = \frac{P^2}{2} (P_T^{\mu\nu} + P_L^{\mu\nu}) - \ak^2 \sin^2\theta^\ast_\pm P_T^{\mu\nu} -  \frac{P^2}{2  \aP^2} (P_0 \pm 2 \omega )^2 P_L^{\mu\nu}.
\end{align}
The final result then reads
\begin{equation}
\label{eq:Pi12fullresult}
  \Pi^{\mu\nu}_{12} = -i \frac{2\alphaEM}{\aP} \int_{m_e}^\infty \dd{\omega} \sum_{\pm} \Theta(1- |\cos\theta^\ast_\pm|)  (\fFermi(\omega )-\Theta (k^0))  (\fFermi(| k^0+P^0| )-\Theta (-k^0-P^0))   T_\pm^{\mu\nu},
\end{equation}
where the sum runs over positive and negative energies $k^0 = \pm \omega$. 
In the HTL limit we find
\begin{align}
\label{eq:Pi12LHTL}
    &\Pi^L_{12} \HTLeq -i\frac{\pi e^2 P^2 T^3}{3 \aP^3} \Theta(\aP - |P^0|),\\
    &\Pi^T_{12} \HTLeq i\frac{\pi e^2 P^2 T^3}{6 \aP^3} \Theta(\aP - |P^0|),
\label{eq:Pi12THTL}
\end{align}
where we have used the integral $\int_0^\infty \dd{k} \fFermi(k) \left[1-\fFermi(k)\right]=\pi^2 T^3/6$.

As will be explained in the subsequent section, it is convenient to compute the retarded and advanced photon self-energy $\Pi_{R/A}$ in order to perform the resummation of the photon propagator. Using the relation $\mathrm{Im}\, \Pi_{11} = \frac{i}{2}(\Pi_{12} + \Pi_{21})$, we can write the advanced and retarded self-energies as 
\begin{equation}
\begin{aligned}
\label{eq:ret/advselfenergy}
    \Pi^{T/L}_{R/A} = \Pi^{T/L}_{11} + \Pi^{T/L}_{12/21} &= \mathrm{Re}\, \Pi^{T/L}_{11} \pm \frac{1}{2}\left(\Pi^{T/L}_{12}-\Pi^{T/L}_{21}\right)\\
    &= \mathrm{Re}\, \Pi^{T/L}_{11} \pm \frac{1}{2}\left(1-e^{P_0/T}\right)\Pi^{T/L}_{12},
\end{aligned}
\end{equation}
where we have used the KMS relation \eqref{eq:KMSrelation} at the second equality. In particular, equation \eqref{eq:ret/advselfenergy} implies that $\mathrm{Im}\, \Pi^{R} = -\mathrm{Im}\, \Pi^{A}$. With this in mind, we can directly compare our results for the advanced and the retarded transverse photon self-energies in the HTL limit to the results of Carrington {\it et al.}~\cite{Carrington:1997sq}.
Given that $\Pi^{L/T}_{R,A} \HTLeq \mathrm{Re} \, \Pi^{L/T}_{11}\mp \frac{P^0}{2T}  \Pi^{L/T}_{12}$ in the HTL limit, our results are in agreement with~\cite{Carrington:1997sq} after carefully sending the time-ordering parameter $\epsilon$ to zero.
In the longitudinal part we differ by a term $P_0^2/\aP^2$ with respect to~\cite{Carrington:1997sq},%
\footnote{This difference can be seen by comparing figure~\ref{fig:ComparisonHTLvsfullself-energy_PiR_differentT/me} to figure 4 in \cite{Peshier:1998dy}.} 
but agree with \cite{Bellac:2011kqa}. 
We note in passing that the self-energy \eqref{eq:Pi12fullresult} leads to the unphysical process of photon decay $\gamma \to e^+e^-$ at high enough temperatures, where $m_\gamma$ exceeds $2m_e$~\cite{Braaten:1993jw}. In practice, this could be resolved by resumming the electron propagator. In our case, however, the relevant dynamics happen at temperatures much below that threshold, such that this resummation is not necessary.

\begin{figure}
    \centering
    \includegraphics[width=.495\textwidth]{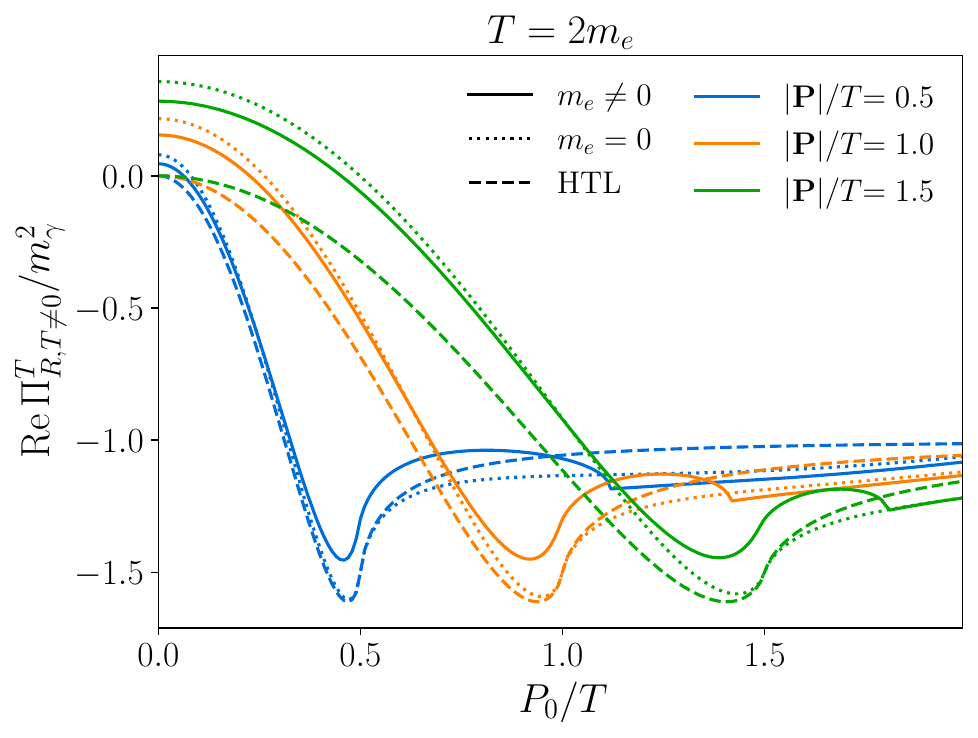}
    \includegraphics[width=.495\textwidth]{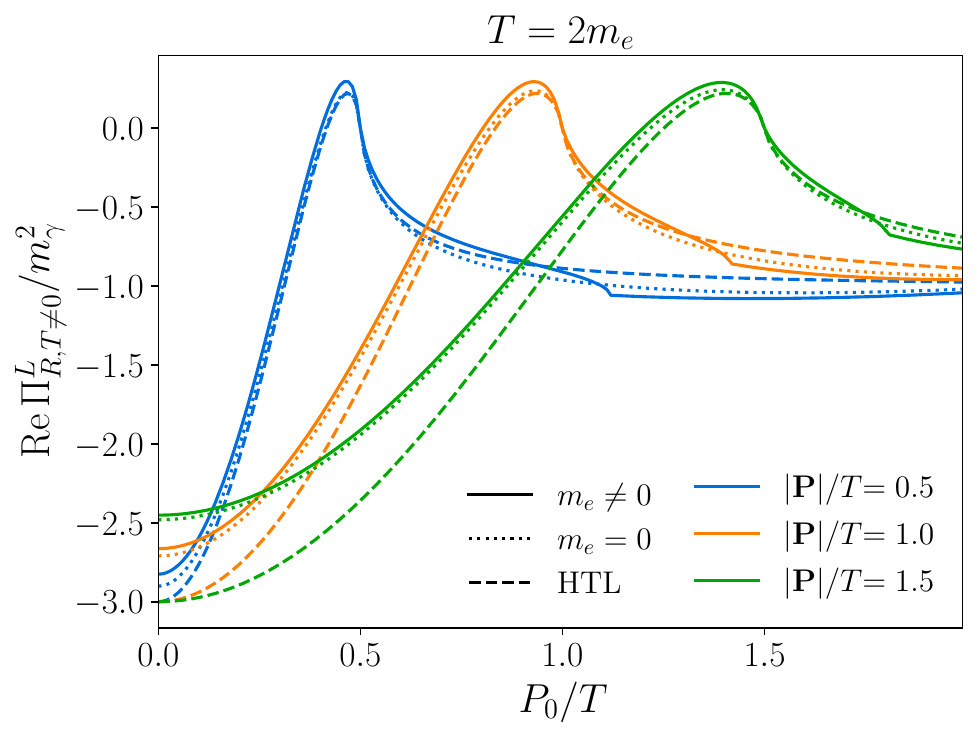}
    \includegraphics[width=.495\textwidth]{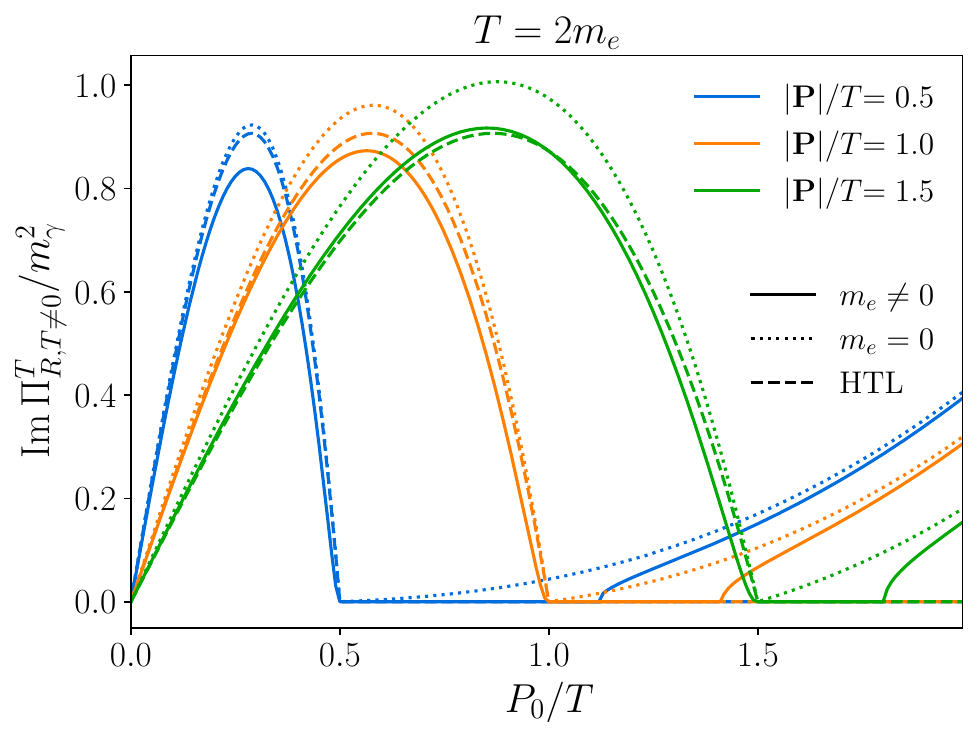}
    \includegraphics[width=.495\textwidth]{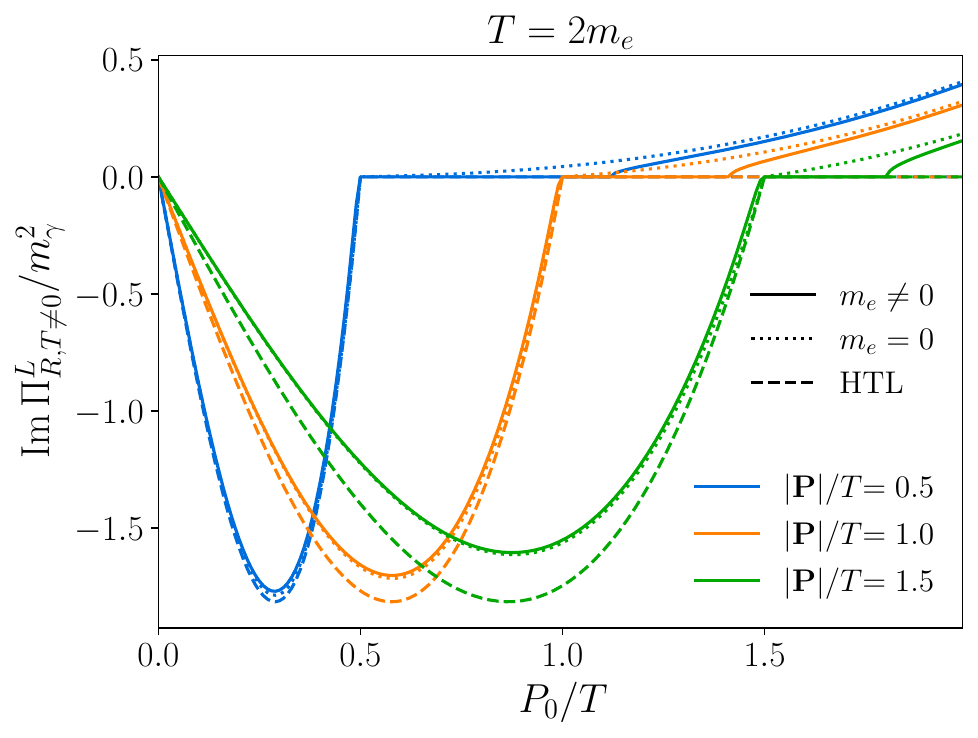}
    \caption{Comparison between the finite-temperature part of the photon self-energy at the one-loop level with (continuous lines) and without (dashed lines) the HTL approximation at $T=2 m_e$ and for various choices of $\aP/T$.}
    \label{fig:ComparisonHTLvsfullself-energy_PiR}
\end{figure}

\begin{figure}
    \centering
    \includegraphics[width=.495\textwidth]{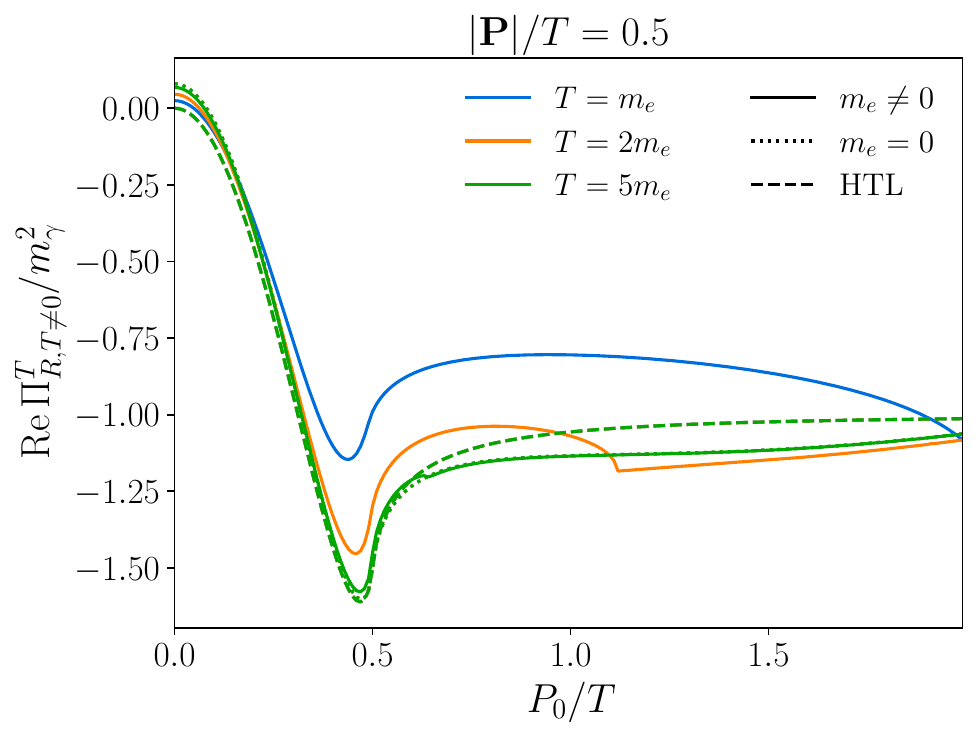}
    \includegraphics[width=.495\textwidth]{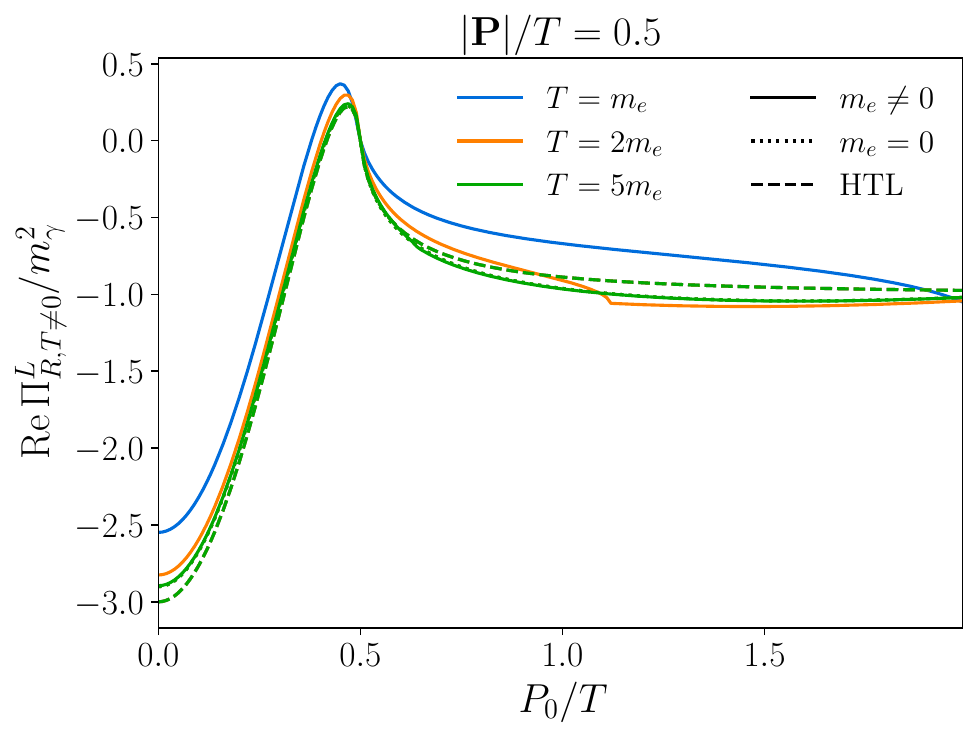}
    \caption{Comparison between the finite-temperature part of the photon self-energy at the one-loop level with (continuous lines) and without (dashed lines) the HTL approximation for different temperatures $T \in \{m_e,2 m_e,5 m_e\}$ at a fixed $\aP/T=0.5$.}
    \label{fig:ComparisonHTLvsfullself-energy_PiR_differentT/me}
\end{figure}

In figure~\ref{fig:ComparisonHTLvsfullself-energy_PiR}, we display the finite-temperature contribution to the real and the imaginary parts of the retarded transverse and longitudinal propagators, for $T=2m_e$ and various choices of~$\aP/T$. We compare the exact one-loop photon self-energy to (i)~the equivalent self-energy in the limit $m_e\rightarrow 0$, and (ii)~the equivalent self-energy in the HTL limit. Similarly, in figure~\ref{fig:ComparisonHTLvsfullself-energy_PiR_differentT/me}, we examine the impact of the temperature on the photon self-energy. We observe that the effect of the finite electron mass~$m_e$, while small for large temperatures $T/m_e \sim 5$, remains sizeable for temperatures around the electron neutrino decoupling temperature where $T/m_e \sim 2$.

\begin{figure}
    \centering
    \includegraphics[width=.6\textwidth]{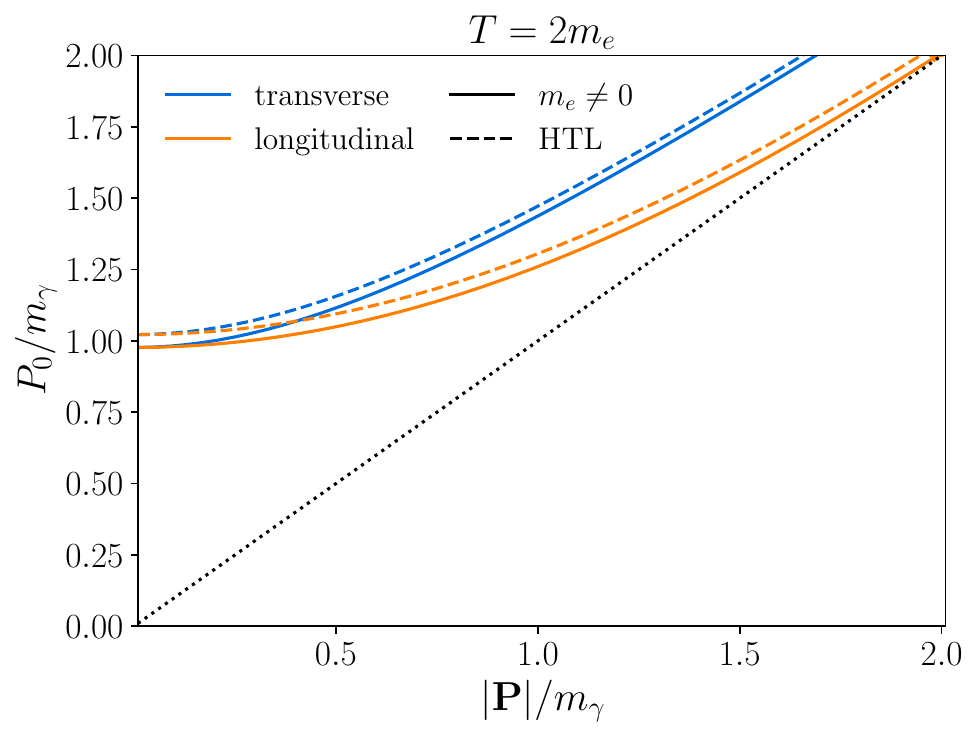}
    \caption{Dispersion relation for transverse (blue lines) and longitudinal (orange lines) photons at the one-loop level with (continuous lines) and without (dashed lines) the HTL approximation at $T=2 m_e$. The black dotted line indicates the dispersion relation at zero temperature, i.e., $P_0^2 = |\mathbf{P}|^2$.}
    \label{fig:Dispersionrelation}
\end{figure}

Beyond the photon self-energy, other quantities of interest are the residue of the transverse and longitudinal photon%
\footnote{Longitudinal photons are also sometimes referred to as ``plasmons'' in the literature.} 
propagators which are quantified by 
\begin{align}
    Z^{T/L} = \lim_{P_0\rightarrow w(|\mathbf{P}|)}  \left[1+\frac{\partial \mathrm{Re}\, \Pi^{T/L}_{11}}{\partial P_0^2}\right]^{-1},
\end{align}
where $P_0 = w(|\mathbf{P}|)$ defines the photon dispersion relation, i.e., solutions of the equation $P^2+\mathrm{Re}\, \Pi^{T/L}_{11}(P) = 0$, which we show for the benchmark $T=2m_e$ case in figure~\ref{fig:Dispersionrelation}. The two residues take the form
\begin{equation}
\begin{aligned}
    Z^L = & \Bigg[1-\frac{\alphaEM}{\pi |\mathbf{P}|^3} \int_{m_e}^\infty \d \omega \fFermi(\omega) \Big[\ell'_2(\omega,P) \left(4\omega P_0P^2\right) +\ell_2(\omega,P) \left(2\omega(3P_0-|\mathbf{P}|^2/P_0)\right)\\
    &+\ell'_1(\omega,P) \left(P^4+4\omega^2P^2\right)+\ell_1(\omega,P) \left(2P^2+4\omega^2\right)-8\ak|\mathbf{P}| \Big]_{\mid P_0 = w(|\mathbf{P}|)}\Bigg]^{-1},
\end{aligned}
\end{equation}
and
\begin{equation}
\begin{aligned}
    Z^T =& \Bigg[1+\frac{\alphaEM}{2\pi |\mathbf{P}|^3} \int_{m_e}^\infty \d \omega \fFermi(\omega) \Big[\ell'_2(\omega,P) \left(4\omega P_0P^2\right) +\ell_2(\omega,P) \left(2\omega(3P_0-|\mathbf{P}|^2/P_0)\right)\\
    &+\ell'_1(\omega,P) \left(P^4+4\omega^2P_0^2-4\ak^2|\mathbf{P}|^2+2P^2|\mathbf{P}|^2\right)\\
    &+\ell_1(\omega,P) \left(4\omega^2+2P_0^2\right)-8\ak|\mathbf{P}| \Big]_{\mid P_0 = w(|\mathbf{P}|)}\Bigg]^{-1},
\end{aligned}
\end{equation}
where we have defined the following derivatives
\begin{equation}
\begin{aligned}
    \ell'_1(\omega,P) \equiv \frac{\partial \ell_1}{\partial P_0^2} &= \frac{2(P^2 + 2 \ak |\mathbf{P}|) - 4  \omega^2}{(P^2 + 2 \ak |\mathbf{P}|)^2 - 4 P_0^2 \omega^2}-\frac{2(P^2 - 2 \ak |\mathbf{P}|) - 4  \omega^2}{(P^2 - 2 \ak |\mathbf{P}|)^2 - 4 P_0^2 \omega^2}, \\
    \ell'_2(\omega,P) \equiv \frac{\partial \ell_2}{\partial P_0^2} &= \frac{2P^2-8\omega( P_0 \omega +  |\mathbf{P}\ak)}{P^4  - 4 ( P_0 \omega +  |\mathbf{P}| \ak)^2}-\frac{2P^2-8\omega( P_0 \omega -  |\mathbf{P}| \ak)}{P^4  - 4 ( P_0 \omega -  |\mathbf{P}| \ak)^2},
\end{aligned}
\end{equation}
of the logarithmic functions  $\ell_1(\omega, P)$ and  $\ell_2(\omega, P)$ of equation~\eqref{eq:logfunc}.

\subsection{The resummed photon propagator}

 The resummed photon propagator $\bar{D}^{ab}_{\mu\nu}$ is given by the Dyson-Schwinger equation
\begin{equation}
  \bar{D}^{ab}_{\mu\nu} = D^{ab}_{\mu\nu} + \sum_{c,d}  D^{ac}_{\mu\alpha} (\Pi^{\alpha\beta})^{cd} \bar{D}^{db}_{\beta\nu}.
\end{equation}
The sum over the real-time labels $c$ and $d$ can be avoided \cite{Ghiglieri:2020dpq} by re-expressing the equation in terms of the retarded and advanced propagators 
\begin{equation}
\begin{aligned}
  &G^R = G_{11} - G_{12} = G_{21} - G_{22}, \\
  &G^A = G_{11} - G_{21} = G_{12} - G_{22},  
\end{aligned}
\end{equation}
defined in a similar fashion to the retarded/advanced self-energies~\eqref{eq:ret/advselfenergy}. This leads to the compact result
\begin{equation}
  \bar{D}^{R/A}_{\mu\nu} = D^{R/A}_{\mu\nu} + D^{R/A}_{\mu\alpha} (\Pi^{\alpha\beta})^{R/A} \bar{D}^{R/A}_{\beta\nu}.
  \label{eq:DysonSchwinger_RA}
\end{equation}
Decomposing the resummed photon propagator using the projectors from equations \eqref{eq:projectors},
\begin{equation}
\bar{D}^{ab}_{\mu\nu} = d_{\mu\nu} \bar{D}^{ab}= -\bar{D}_L^{ab} P^{L}_{\mu\nu} -  \bar{D}_T^{ab} P^{T}_{\mu\nu} - \xi \Delta^{ab} \frac{P_\mu P_\nu}{P^2},
\end{equation}
and using equation \eqref{eq:DysonSchwinger_RA}, the corresponding longitudinal and transverse components of the resummed retarded and advanced propagator take the form
\begin{equation}
  \bar{D}_{L/T}^{R/A} = \frac{1}{(\Delta^{R/A})^{-1}  + \Pi^{R/A}_{L/T}}.
\end{equation}
With $D_{11} = \frac{1}{2}(D^A + D^R + 2 D^+)$ and using the fact that \emph{in equilibrium}, the thermal resummed statistical photon propagator $\bar{D}_{\mu\nu}^+$ is related to the retarded and advanced propagators via~\cite{Carrington:1997sq}
\begin{equation}
  \bar{D}_{\mu\nu}^+(P) = \frac{1}{2}[1 + 2 f_B(|P^0|)] \mathrm{sgn}(P^0) \left[\bar{D}^R_{\mu\nu} - \bar{D}_{\mu\nu}^A\right],
\end{equation}
we finally arrive at
\begin{align}
    \bar{D}^{L/T}_{11} = \frac{\Omega_{L,T}}{\Omega^2_{L,T} +\Gamma^2_{L,T} } - i [1 + 2 f_B(|P^0|)] \mathrm{sgn}(P^0) \frac{\Gamma_{L,T}}{\Omega^2_{L,T} +\Gamma^2_{L,T} } ,
    \label{eq:D11resummed}
\end{align}
with 
\begin{align}
    \Omega^{L,T} &= P^2 + \mathrm{Re}\, \Pi^{L/T}_R, \\
    \Gamma^{L,T} &=  \mathrm{Im}\, \Pi^{L,T}_R = -\frac{1}{2f_B(P^0)}  \mathrm{Im} \Pi^{L,T}_{12}.\label{GammaDefPhoton}
\end{align}
It is understood that \eqref{GammaDefPhoton} should be replaced by the causality-respecting $\epsilon$-prescription if the imaginary part of the photon self-energy vanishes for kinematic reasons at the given order in perturbation theory.


\section{Parameterisation of the collision integrals}
\label{app:param}
We parameterise the momenta in the self-energy expression~\eqref{eq:amp_simple} as in reference~\cite{Bennett:2019ewm},
\begin{equation}
  \begin{aligned}
    &\mathbf{p} = \ap (0,0,1)^T,\\
    &\mathbf{l} = \al (0,\sin\alpha,\cos\alpha)^T, \\
    &\mathbf{q} = \aq (\sin\theta\sin\beta,\sin\theta\cos\beta,\cos\theta)^T,
  \end{aligned}
  \end{equation}
resulting in 
\begin{equation}
  \begin{aligned}
    &\mathrm{Tr}\left[\slashed{p}\Sigma^{12}_{\mathrm{NLO}}\right]=\int  \dd[4]{q} \dd[4]{l}\delta(l^2)\delta(q^2 - m_e^2) \delta((q+p-l)^2 - m_e^2) f(q,l,p) \\
    &= \left. \int \frac{\dd[3]{\mathbf{q}} \dd[3]{\mathbf{l}}}{4 \al E_q} \sum_{\epsilon,\tau=\pm 1} f(q,l,p) \delta((q+p-l)^2 - m_e^2) \right|_{\overset{q^0=\epsilon E_q}{l^0 = \tau \al}} \\
    &= \frac{2\pi}{4}\left.  \int \dd{\al}   \dd{\aq} \frac{\aq^2 \al}{E_q} \int_{-1}^1 \dd{\!\cos\alpha} \sum_{\epsilon,\tau=\pm 1}\frac{\theta(b^2 - 4 a c)}{\sqrt{|a|}} \int_{z_-}^{z^+} \dd{z}  \frac{f(q,l,p)}{\sqrt{(z - z_{-})(z_+ - z)}}\right|_{\overset{q^0=\epsilon E_q}{l^0 = \tau \al}},
  \end{aligned}
  \end{equation}
  where $E_q = \sqrt{|\mathbf{q}|^2 + m_e^2}$, and the function $f(q,l,p)$ acts as a placeholder for the integrand in equation \eqref{eq:amp_simple}.
  The last $\delta$-function fixes the angle $\beta$ to
  \begin{equation}
    \cos\beta_i = \frac{\ap \aq \cos\theta + \epsilon \tau E_q \al + \tau \al \ap - \ap \al \cos\alpha - \epsilon E_q \ap - \al \aq \cos\alpha \cos\theta}{\al \aq \sin\alpha \sin\theta}.
  \end{equation}
 The integration limits on $z=\cos\theta$,
  \begin{equation}
    z_\pm = -\frac{b}{2a}\pm \sqrt{\left(\frac{b}{2a}\right)^2 - \frac{c}{a}},
  \end{equation}
are the roots of the quadratic polynomial $a z^2 + b z + c$, with 
  \begin{equation}
  \begin{aligned}
    a &= -q^2 |\mathbf{p} - \mathbf{l}|^2, \\
    b &= -2 (\cos\alpha |\mathbf{l}| - |\mathbf{p}|) |\mathbf{q}| (|\mathbf{l}| |\mathbf{p}| (\cos\alpha - \tau) + \epsilon E_q (|\mathbf{p}| - |\mathbf{l}| \tau)), \\
    c &= (E_q^2 (|\mathbf{p}| - |\mathbf{l}|\tau)^2) + 2 \epsilon E_q |\mathbf{l}| |\mathbf{p}| (\cos\alpha - \tau) (-|\mathbf{p}| + |\mathbf{l}| \tau) \\ 
    &\hspace{2cm} + |\mathbf{l}|^2 (-((\cos^2\alpha) |\mathbf{p}|^2) +  |\mathbf{q}|^2 (1 - \cos^2\alpha) + 2 \cos\alpha |\mathbf{p}|^2 \tau - |\mathbf{p}|^2) 
  \end{aligned}
  \end{equation}
as the coefficients.

The $z$-integration of the vacuum and the thermal corrections can be performed analytically following
  \begin{equation}
      \int_{z_-}^{z^+} \dd{z}  \frac{\mathcal{T}(q,l,p)}{\sqrt{(z - z_{-})(z_+ - z)}} = \mathcal{A}\,  \pi  \left(\mathcal{G}^0 -\frac{b}{2 a}\mathcal{G}^1 + \frac{3 b^2 - 4 c a}{8 a^2}\mathcal{G}^2\right),
  \end{equation}
where we have expanded the correction matrix ${\cal T}$ in powers of $z$ according to 
   \begin{equation}
       \mathcal{T} = \mathcal{A}  \left(\mathcal{G}^0 + \mathcal{G}^1 z + \mathcal{G}^2 z^2\right).
   \end{equation}
Then, for the transverse thermal part, the coefficients read
 \begin{equation}
\begin{aligned} 
    &\mathcal{G}^0_{\mathrm{th},T} = \frac{P^2}{4 \aP^2} \Bigl[ 2 \al^2 P^2+4 l_0 \ap P_0  q_0+8 l_0 \ap
    q_0^2+2 l_0 P_0 P^2+2 m_e^2 
   \aP^2+8 \ap^2 P_0^2 q_0^2/P^2   \\ 
   &  \ \ \ \ \ \ \ \ \ \ -8 \ap^2 q_0^2+4 \ap P_0^2  q_0+4 \ap P_0  q_0^2-2
   \ap P^2 q_0+P_0^2 P^2+2 P_0 P^2 q_0+2
   P^2 q_0^2\Bigr], \\
   &\mathcal{G}^1_{\mathrm{th},T} = -\frac{ \ap \aq P^2}{2 \aP^2}  \left[2 l_0  (P_0+2 q_0)+2 P_0^2 (1 +4 \ap
   q_0/P^2)-8 \ap q_0-P^2+2 P_0 
   q_0\right],\\
    &\mathcal{G}^2_{\mathrm{th},T} = 2 \ap^2 \aq^2,
\end{aligned}
\end{equation}
while for the longitudinal thermal part we have
\begin{equation}
\begin{aligned}
      &\mathcal{G}^0_{\mathrm{th},L} = -\frac{P^2}{8 \aP^2}  \Bigl[4 P_0 q_0 (P^2 + 2\ap q_0) + 4P^2 l^2 + (P^2 + 4 \ap q_0) P^2_0 + P^2(P^2 + 4 q^2_0) \\
       &  \ \ \ \ \ \ \ \ \ \ + 4 l_0 (P_0 (P^2 + 2 \ap q_0) + 4 \ap q^2_0) \Bigr], \\
     &\mathcal{G}^1_{\mathrm{th},L} = \frac{P^2 \aq \ap}{2 \aP^2} (l_0 + \ap) (P_0 + 2q_0),\\
    &\mathcal{G}^2_{\mathrm{th},L} = 0,
\end{aligned}
\end{equation}
with the prefactor $\mathcal{A}^{L/T}_{\rm th} =  -2^8 G_F^2 (g_V^\alpha)^2 \mathrm{Re}\, \bar{D}^{L,T}_{11}  \mathrm{Re}\,\Pi^{L,T}_{11,T\neq 0}$. For the vacuum part, the corresponding coefficients are
\begin{equation}
\begin{aligned}
    &\mathcal{G}^0_{\mathrm{vac}} =  - m_e^2 (l\cdot p)+ 2 \ap^2 q_0^2, \\
   &\mathcal{G}^1_{\mathrm{vac}} = -4 \aq q_0 \ap^2, \\
    &\mathcal{G}^2_{\mathrm{vac}} = 2 \ap^2 \aq^2,
\end{aligned}
\end{equation}
with $\mathcal{A}_{\rm vac} =  -2^{10} G_F^2\, \alphaEM \, [(g^\alpha_V)^2/(4 \pi)]  \, \Pi_2 $.


\section{Functions in the continuity equation}
\label{sec:continuitysupplement}

The continuity~\eqref{eq:dz/dx}, written in terms of the rescaled variables $x,y,z$, contains the functions $J(x/z)$, $Y(x/z)$, and $G_{1,2}(x/z)$.  We give the explicit forms of these functions here to ${\cal O}(e^2)$~\cite{Mangano:2001iu},
\begin{equation}
\begin{aligned}
J(\tau) & \equiv \frac{1}{\pi^2} \int_0^\infty {\rm d}\omega \, \omega^2 \frac{\exp(\sqrt{\omega^2 + \tau^2})}{[\exp(\sqrt{\omega^2 + \tau^2})+1]^2},\\
Y(\tau) & \equiv \frac{1}{\pi^2} \int_0^\infty {\rm d}\omega \, \omega^4 \frac{\exp(\sqrt{\omega^2 + \tau^2})}{[\exp(\sqrt{\omega^2 + \tau^2})+1]^2},
    \end{aligned}    
\end{equation}
\begin{equation}
\begin{aligned}
G_1(\tau) =& 2 \pi \alpha_{\rm em} \Bigg[\frac{1}{\tau}
\left(\frac{K(\tau)}{3} + 2 K^2(\tau)- \frac{J(\tau)}{6}- K(\tau)J(\tau)\right) \\
&+ \left(\frac{K'(\tau)}{6}- K(\tau) K'(\tau) + \frac{J'(\tau)}{6} + J'(\tau) K(\tau) + J(\tau) K'(\tau)\right)\Bigg],
\end{aligned}
\end{equation}
\begin{equation}
\begin{aligned}
G_2(\tau) =& - 8 \pi \alpha_{\rm em} 
\left(\frac{K(\tau)}{6} +\frac{J(\tau)}{6} - \frac{1}{2} K^2(\tau)+K(\tau)J(\tau)\right) \\
&+ 2 \pi \alpha_{\rm em}\left(\frac{K'(\tau)}{6}- K(\tau) K'(\tau) + \frac{J'(\tau)}{6} + J'(\tau) K(\tau) + J(\tau) K'(\tau)\right),
\end{aligned}
\end{equation}
where $'$ denotes a derivative with respect to $\tau$.
Expressions for $G_{1,2}(\tau)$ at ${\cal O}(e^3)$ can be found in reference~\cite{Bennett:2019ewm}.


\bibliography{references.bib}

\providecommand{\href}[2]{#2}\begingroup\raggedright\begin{thebibliography}{10}

\bibitem{Steigman:1977kc}
G.~Steigman, D.N.~Schramm and J.E.~Gunn, \emph{{Cosmological Limits to the
  Number of Massive Leptons}},
  \href{https://doi.org/10.1016/0370-2693(77)90176-9}{\emph{Phys. Lett. B}
  {\bfseries 66} (1977) 202}.

\bibitem{Barbieri:1990vx}
R.~Barbieri and A.~Dolgov, \emph{{Neutrino oscillations in the early
  universe}}, \href{https://doi.org/10.1016/0550-3213(91)90396-F}{\emph{Nucl.
  Phys. B} {\bfseries 349} (1991) 743}.

\bibitem{Abdullahi:2022jlv}
A.M.~Abdullahi et~al., \emph{{The present and future status of heavy neutral
  leptons}}, \href{https://doi.org/10.1088/1361-6471/ac98f9}{\emph{J. Phys. G}
  {\bfseries 50} (2023) 020501}
  [\href{https://arxiv.org/abs/2203.08039}{{\ttfamily 2203.08039}}].

\bibitem{DiLuzio:2022gsc}
L.~Di~Luzio, J.~Martin~Camalich, G.~Martinelli, J.A.~Oller and G.~Piazza,
  \emph{{Axion-pion thermalization rate in unitarized NLO chiral perturbation
  theory}}, \href{https://doi.org/10.1103/PhysRevD.108.035025}{\emph{Phys. Rev.
  D} {\bfseries 108} (2023) 035025}
  [\href{https://arxiv.org/abs/2211.05073}{{\ttfamily 2211.05073}}].

\bibitem{DEramo:2021lgb}
F.~D'Eramo, F.~Hajkarim and S.~Yun, \emph{{Thermal QCD Axions across
  Thresholds}}, \href{https://doi.org/10.1007/JHEP10(2021)224}{\emph{JHEP}
  {\bfseries 10} (2021) 224}
  [\href{https://arxiv.org/abs/2108.05371}{{\ttfamily 2108.05371}}].

\bibitem{Caprini:2018mtu}
C.~Caprini and D.G.~Figueroa, \emph{{Cosmological Backgrounds of Gravitational
  Waves}}, \href{https://doi.org/10.1088/1361-6382/aac608}{\emph{Class. Quant.
  Grav.} {\bfseries 35} (2018) 163001}
  [\href{https://arxiv.org/abs/1801.04268}{{\ttfamily 1801.04268}}].

\bibitem{Aboubrahim:2022gjb}
A.~Aboubrahim, M.~Klasen and P.~Nath, \emph{{Analyzing the Hubble tension
  through hidden sector dynamics in the early universe}},
  \href{https://doi.org/10.1088/1475-7516/2022/04/042}{\emph{JCAP} {\bfseries
  04} (2022) 042} [\href{https://arxiv.org/abs/2202.04453}{{\ttfamily
  2202.04453}}].

\bibitem{Agrawal:2021dbo}
P.~Agrawal et~al., \emph{{Feebly-interacting particles: FIPs 2020 workshop
  report}}, \href{https://doi.org/10.1140/epjc/s10052-021-09703-7}{\emph{Eur.
  Phys. J. C} {\bfseries 81} (2021) 1015}
  [\href{https://arxiv.org/abs/2102.12143}{{\ttfamily 2102.12143}}].

\bibitem{Aghanim:2018eyx}
{\scshape Planck} collaboration, \emph{{Planck 2018 results. VI. Cosmological
  parameters}},
  \href{https://doi.org/10.1051/0004-6361/201833910}{\emph{Astron.\ Astrophys.}
  {\bfseries 641} (2020) A6}
  [\href{https://arxiv.org/abs/1807.06209}{{\ttfamily 1807.06209}}].

\bibitem{Dodelson:1992km}
S.~Dodelson and M.S.~Turner, \emph{{Nonequilibrium neutrino statistical
  mechanics in the expanding universe}},
  \href{https://doi.org/10.1103/PhysRevD.46.3372}{\emph{Phys.\ Rev.\ D}
  {\bfseries 46} (1992) 3372}.

\bibitem{Hannestad:1995rs}
S.~Hannestad and J.~Madsen, \emph{{Neutrino decoupling in the early universe}},
  \href{https://doi.org/10.1103/PhysRevD.52.1764}{\emph{Phys.\ Rev.\ D}
  {\bfseries 52} (1995) 1764}
  [\href{https://arxiv.org/abs/astro-ph/9506015}{{\ttfamily
  astro-ph/9506015}}].

\bibitem{Dolgov:1997mb}
A.D.~Dolgov, S.H.~Hansen and D.V.~Semikoz, \emph{{Nonequilibrium corrections to
  the spectra of massless neutrinos in the early universe}},
  \href{https://doi.org/10.1016/S0550-3213(97)00479-3}{\emph{Nucl.\ Phys.\ B}
  {\bfseries 503} (1997) 426}.

\bibitem{Dolgov:1998sf}
A.D.~Dolgov, S.H.~Hansen and D.V.~Semikoz, \emph{{Nonequilibrium corrections to
  the spectra of massless neutrinos in the early universe: Addendum}},
  \href{https://doi.org/10.1016/S0550-3213(98)00818-9}{\emph{Nucl.\ Phys.\ B}
  {\bfseries 543} (1999) 269}.

\bibitem{Esposito:2000hi}
S.~Esposito, G.~Miele, S.~Pastor, M.~Peloso and O.~Pisanti,
  \emph{{Nonequilibrium spectra of degenerate relic neutrinos}},
  \href{https://doi.org/10.1016/S0550-3213(00)00554-X}{\emph{Nucl.\ Phys.\ B}
  {\bfseries 590} (2000) 539}.

\bibitem{Froustey:2019owm}
J.~Froustey and C.~Pitrou, \emph{{Incomplete neutrino decoupling effect on big
  bang nucleosynthesis}},
  \href{https://doi.org/10.1103/PhysRevD.101.043524}{\emph{Phys.\ Rev.\ D}
  {\bfseries 101} (2020) 043524}
  [\href{https://arxiv.org/abs/1912.09378}{{\ttfamily 1912.09378}}].

\bibitem{Dicus:1982bz}
D.A.~Dicus, E.W.~Kolb, A.M.~Gleeson, E.C.G.~Sudarshan, V.L.~Teplitz and
  M.S.~Turner, \emph{{Primordial Nucleosynthesis Including Radiative, Coulomb,
  and Finite Temperature Corrections to Weak Rates}},
  \href{https://doi.org/10.1103/PhysRevD.26.2694}{\emph{Phys.\ Rev.\ D}
  {\bfseries 26} (1982) 2694}.

\bibitem{Heckler:1994tv}
A.~Heckler, \emph{{Astrophysical applications of quantum corrections to the
  equation of state of a plasma}},
  \href{https://doi.org/10.1103/PhysRevD.49.611}{\emph{Phys.\ Rev.\ D}
  {\bfseries 49} (1994) 611}.

\bibitem{Fornengo:1997wa}
N.~Fornengo, C.~Kim and J.~Song, \emph{{Finite temperature effects on the
  neutrino decoupling in the early universe}},
  \href{https://doi.org/10.1103/PhysRevD.56.5123}{\emph{Phys.\ Rev.\ D}
  {\bfseries 56} (1997) 5123}
  [\href{https://arxiv.org/abs/hep-ph/9702324}{{\ttfamily hep-ph/9702324}}].

\bibitem{Lopez:1998vk}
R.E.~Lopez and M.S.~Turner, \emph{{An Accurate Calculation of the Big Bang
  Prediction for the Abundance of Primordial Helium}},
  \href{https://doi.org/10.1103/PhysRevD.59.103502}{\emph{Phys.\ Rev.\ D}
  {\bfseries 59} (1999) 103502}
  [\href{https://arxiv.org/abs/astro-ph/9807279}{{\ttfamily
  astro-ph/9807279}}].

\bibitem{Mangano:2001iu}
G.~Mangano, G.~Miele, S.~Pastor and M.~Peloso, \emph{{A Precision calculation
  of the effective number of cosmological neutrinos}},
  \href{https://doi.org/10.1016/S0370-2693(02)01622-2}{\emph{Phys.\ Lett.\ B}
  {\bfseries 534} (2002) 8}
  [\href{https://arxiv.org/abs/astro-ph/0111408}{{\ttfamily
  astro-ph/0111408}}].

\bibitem{Bennett:2019ewm}
J.J.~Bennett, G.~Buldgen, M.~Drewes and Y.Y.Y.~Wong, \emph{{Towards a precision
  calculation of the effective number of neutrinos $N_{\rm eff}$ in the
  Standard Model I: the QED equation of state}},
  \href{https://doi.org/10.1088/1475-7516/2020/03/003}{\emph{JCAP} {\bfseries
  03} (2020) 003} [\href{https://arxiv.org/abs/1911.04504}{{\ttfamily
  1911.04504}}].

\bibitem{Mangano:2005cc}
G.~Mangano, G.~Miele, S.~Pastor, T.~Pinto, O.~Pisanti and P.D.~Serpico,
  \emph{{Relic neutrino decoupling including flavor oscillations}},
  \href{https://doi.org/10.1016/j.nuclphysb.2005.09.041}{\emph{Nucl.\ Phys.\ B}
  {\bfseries 729} (2005) 221}
  [\href{https://arxiv.org/abs/hep-ph/0506164}{{\ttfamily hep-ph/0506164}}].

\bibitem{Birrell:2014uka}
J.~Birrell, C.-T.~Yang and J.~Rafelski, \emph{{Relic Neutrino Freeze-out:
  Dependence on Natural Constants}},
  \href{https://doi.org/10.1016/j.nuclphysb.2014.11.020}{\emph{Nucl.\ Phys.\ B}
  {\bfseries 890} (2014) 481}
  [\href{https://arxiv.org/abs/1406.1759}{{\ttfamily 1406.1759}}].

\bibitem{Grohs2016}
E.~Grohs, G.M.~Fuller, C.T.~Kishimoto, M.W.~Paris and A.~Vlasenko,
  \emph{{Neutrino energy transport in weak decoupling and big bang
  nucleosynthesis}},
  \href{https://doi.org/10.1103/PhysRevD.93.083522}{\emph{Phys.\ Rev.\ D}
  {\bfseries 93} (2016) 083522}
  [\href{https://arxiv.org/abs/1512.02205}{{\ttfamily 1512.02205}}].

\bibitem{deSalas:2016ztq}
P.F.~de~Salas and S.~Pastor, \emph{{Relic neutrino decoupling with flavour
  oscillations revisited}},
  \href{https://doi.org/10.1088/1475-7516/2016/07/051}{\emph{JCAP} {\bfseries
  07} (2016) 051} [\href{https://arxiv.org/abs/1606.06986}{{\ttfamily
  1606.06986}}].

\bibitem{Gariazzo:2019gyi}
S.~Gariazzo, P.F.~de~Salas and S.~Pastor, \emph{{Thermalisation of sterile
  neutrinos in the early Universe in the 3+1 scheme with full mixing matrix}},
  \href{https://doi.org/10.1088/1475-7516/2019/07/014}{\emph{JCAP} {\bfseries
  07} (2019) 014} [\href{https://arxiv.org/abs/1905.11290}{{\ttfamily
  1905.11290}}].

\bibitem{Akita:2020szl}
K.~Akita and M.~Yamaguchi, \emph{{A precision calculation of relic neutrino
  decoupling}},
  \href{https://doi.org/10.1088/1475-7516/2020/08/012}{\emph{JCAP} {\bfseries
  08} (2020) 012} [\href{https://arxiv.org/abs/2005.07047}{{\ttfamily
  2005.07047}}].

\bibitem{Froustey:2020mcq}
J.~Froustey, C.~Pitrou and M.C.~Volpe, \emph{{Neutrino decoupling including
  flavour oscillations and primordial nucleosynthesis}},
  \href{https://doi.org/10.1088/1475-7516/2020/12/015}{\emph{JCAP} {\bfseries
  12} (2020) 015} [\href{https://arxiv.org/abs/2008.01074}{{\ttfamily
  2008.01074}}].

\bibitem{Bennett:2020zkv}
J.J.~Bennett, G.~Buldgen, P.F.~De~Salas, M.~Drewes, S.~Gariazzo, S.~Pastor
  et~al., \emph{{Towards a precision calculation of $N_{\rm eff}$ in the
  Standard Model II: Neutrino decoupling in the presence of flavour
  oscillations and finite-temperature QED}},
  \href{https://doi.org/10.1088/1475-7516/2021/04/073}{\emph{JCAP} {\bfseries
  04} (2021) 073} [\href{https://arxiv.org/abs/2012.02726}{{\ttfamily
  2012.02726}}].

\bibitem{Cielo:2023bqp}
M.~Cielo, M.~Escudero, G.~Mangano and O.~Pisanti, \emph{{Neff in the Standard
  Model at NLO is 3.043}},
  \href{https://doi.org/10.1103/PhysRevD.108.L121301}{\emph{Phys. Rev. D}
  {\bfseries 108} (2023) L121301}
  [\href{https://arxiv.org/abs/2306.05460}{{\ttfamily 2306.05460}}].

\bibitem{Esposito:2003wv}
S.~Esposito, G.~Mangano, G.~Miele, I.~Picardi and O.~Pisanti, \emph{{Neutrino
  energy loss rate in a stellar plasma}},
  \href{https://doi.org/10.1016/S0550-3213(03)00151-2}{\emph{Nucl. Phys. B}
  {\bfseries 658} (2003) 217}
  [\href{https://arxiv.org/abs/astro-ph/0301438}{{\ttfamily
  astro-ph/0301438}}].

\bibitem{Jackson:2023zkl}
G.~Jackson and M.~Laine, \emph{{QED corrections to the thermal neutrino
  interaction rate}},  \href{https://arxiv.org/abs/2312.07015}{{\ttfamily
  2312.07015}}.

\bibitem{Tomalak:2019ibg}
O.~Tomalak and R.J.~Hill, \emph{{Theory of elastic neutrino-electron
  scattering}}, \href{https://doi.org/10.1103/PhysRevD.101.033006}{\emph{Phys.
  Rev. D} {\bfseries 101} (2020) 033006}
  [\href{https://arxiv.org/abs/1907.03379}{{\ttfamily 1907.03379}}].

\bibitem{Hill:2019xqk}
R.J.~Hill and O.~Tomalak, \emph{{On the effective theory of neutrino-electron
  and neutrino-quark interactions}},
  \href{https://doi.org/10.1016/j.physletb.2020.135466}{\emph{Phys. Lett. B}
  {\bfseries 805} (2020) 135466}
  [\href{https://arxiv.org/abs/1911.01493}{{\ttfamily 1911.01493}}].

\bibitem{CMB-S4:2016ple}
{\scshape CMB-S4} collaboration, \emph{{CMB-S4 Science Book, First Edition}},
  \href{https://arxiv.org/abs/1610.02743}{{\ttfamily 1610.02743}}.

\bibitem{Landsman:1986uw}
N.P.~Landsman and C.G.~van Weert, \emph{{Real and Imaginary Time Field Theory
  at Finite Temperature and Density}},
  \href{https://doi.org/10.1016/0370-1573(87)90121-9}{\emph{Phys. Rept.}
  {\bfseries 145} (1987) 141}.

\bibitem{Chou:1984es}
K.-c.~Chou, Z.-b.~Su, B.-l.~Hao and L.~Yu, \emph{{Equilibrium and
  Nonequilibrium Formalisms Made Unified}},
  \href{https://doi.org/10.1016/0370-1573(85)90136-X}{\emph{Phys. Rept.}
  {\bfseries 118} (1985) 1}.

\bibitem{Berges:2004yj}
J.~Berges, \emph{{Introduction to nonequilibrium quantum field theory}},
  \href{https://doi.org/10.1063/1.1843591}{\emph{AIP Conf. Proc.} {\bfseries
  739} (2004) 3} [\href{https://arxiv.org/abs/hep-ph/0409233}{{\ttfamily
  hep-ph/0409233}}].

\bibitem{Kapusta:2006pm}
J.I.~Kapusta and C.~Gale, \emph{{Finite-temperature field theory: Principles
  and applications}}, Cambridge Monographs on Mathematical Physics, Cambridge
  University Press (2011),
  \href{https://doi.org/10.1017/CBO9780511535130}{10.1017/CBO9780511535130}.

\bibitem{Coriano:1994re}
C.~Coriano and R.R.~Parwani, \emph{{The Three loop equation of state of QED at
  high temperature}},
  \href{https://doi.org/10.1103/PhysRevLett.73.2398}{\emph{Phys. Rev. Lett.}
  {\bfseries 73} (1994) 2398}
  [\href{https://arxiv.org/abs/hep-ph/9405343}{{\ttfamily hep-ph/9405343}}].

\bibitem{Parwani:1994xi}
R.R.~Parwani and C.~Coriano, \emph{{Higher order corrections to the equation of
  state of QED at high temperature}},
  \href{https://doi.org/10.1016/0550-3213(94)00484-V}{\emph{Nucl. Phys. B}
  {\bfseries 434} (1995) 56}
  [\href{https://arxiv.org/abs/hep-ph/9409269}{{\ttfamily hep-ph/9409269}}].

\bibitem{Sigl:1993ctk}
G.~Sigl and G.~Raffelt, \emph{{General kinetic description of relativistic
  mixed neutrinos}},
  \href{https://doi.org/10.1016/0550-3213(93)90175-O}{\emph{Nucl. Phys. B}
  {\bfseries 406} (1993) 423}.

\bibitem{Canetti:2012zc}
L.~Canetti, M.~Drewes and M.~Shaposhnikov, \emph{{Matter and Antimatter in the
  Universe}}, \href{https://doi.org/10.1088/1367-2630/14/9/095012}{\emph{New J.
  Phys.} {\bfseries 14} (2012) 095012}
  [\href{https://arxiv.org/abs/1204.4186}{{\ttfamily 1204.4186}}].

\bibitem{Antusch:2017pkq}
S.~Antusch, E.~Cazzato, M.~Drewes, O.~Fischer, B.~Garbrecht, D.~Gueter et~al.,
  \emph{{Probing Leptogenesis at Future Colliders}},
  \href{https://doi.org/10.1007/JHEP09(2018)124}{\emph{JHEP} {\bfseries 09}
  (2018) 124} [\href{https://arxiv.org/abs/1710.03744}{{\ttfamily
  1710.03744}}].

\bibitem{Schwinger:1960qe}
J.S.~Schwinger, \emph{{Brownian motion of a quantum oscillator}},
  \href{https://doi.org/10.1063/1.1703727}{\emph{J. Math. Phys.} {\bfseries 2}
  (1961) 407}.

\bibitem{Keldysh:1964ud}
L.V.~Keldysh, \emph{{Diagram technique for nonequilibrium processes}},
  {\emph{Zh. Eksp. Teor. Fiz.} {\bfseries 47} (1964) 1515}.

\bibitem{Garbrecht:2011xw}
B.~Garbrecht and M.~Garny, \emph{{Finite Width in out-of-Equilibrium
  Propagators and Kinetic Theory}},
  \href{https://doi.org/10.1016/j.aop.2011.10.005}{\emph{Annals Phys.}
  {\bfseries 327} (2012) 914}
  [\href{https://arxiv.org/abs/1108.3688}{{\ttfamily 1108.3688}}].

\bibitem{Drewes:2012qw}
M.~Drewes, S.~Mendizabal and C.~Weniger, \emph{{The Boltzmann Equation from
  Quantum Field Theory}},
  \href{https://doi.org/10.1016/j.physletb.2012.11.046}{\emph{Phys. Lett. B}
  {\bfseries 718} (2013) 1119}
  [\href{https://arxiv.org/abs/1202.1301}{{\ttfamily 1202.1301}}].

\bibitem{Braaten:1993jw}
E.~Braaten and D.~Segel, \emph{{Neutrino energy loss from the plasma process at
  all temperatures and densities}},
  \href{https://doi.org/10.1103/PhysRevD.48.1478}{\emph{Phys. Rev. D}
  {\bfseries 48} (1993) 1478}
  [\href{https://arxiv.org/abs/hep-ph/9302213}{{\ttfamily hep-ph/9302213}}].

\bibitem{Kolb:1990vq}
E.W.~Kolb and M.S.~Turner, \emph{{The Early Universe}}, vol.~69 (1990),
  \href{https://doi.org/10.1201/9780429492860}{10.1201/9780429492860}.

\bibitem{Braaten:ApJ}
E.~Braaten, \emph{{Neutrino emissivity of an ultrarelativistic plasma from
  positron and plasmino annihilation}},
  \href{https://doi.org/10.1086/171405}{\emph{Astrophys. J.} {\bfseries 392}
  (1992) 70}.

\bibitem{ParticleDataGroup:2022pth}
{\scshape Particle Data Group} collaboration, \emph{{Review of Particle
  Physics}}, \href{https://doi.org/10.1093/ptep/ptac097}{\emph{PTEP} {\bfseries
  2022} (2022) 083C01}.

\bibitem{Peshier:1998dy}
A.~Peshier, K.~Schertler and M.H.~Thoma, \emph{{One loop selfenergies at finite
  temperature}}, \href{https://doi.org/10.1006/aphy.1997.5781}{\emph{Annals
  Phys.} {\bfseries 266} (1998) 162}
  [\href{https://arxiv.org/abs/hep-ph/9708434}{{\ttfamily hep-ph/9708434}}].

\bibitem{Carrington:1997sq}
M.E.~Carrington, D.-f.~Hou and M.H.~Thoma, \emph{{Equilibrium and
  nonequilibrium hard thermal loop resummation in the real time formalism}},
  \href{https://doi.org/10.1007/s100520050412}{\emph{Eur. Phys. J. C}
  {\bfseries 7} (1999) 347}
  [\href{https://arxiv.org/abs/hep-ph/9708363}{{\ttfamily hep-ph/9708363}}].

\bibitem{Bellac:2011kqa}
M.L.~Bellac, \emph{{Thermal Field Theory}}, Cambridge Monographs on
  Mathematical Physics, Cambridge University Press (3, 2011),
  \href{https://doi.org/10.1017/CBO9780511721700}{10.1017/CBO9780511721700}.

\bibitem{Rebhan:2001iw}
A.~Rebhan, \emph{{Hard thermal loops and QCD thermodynamics}},  in
  \emph{{Cargese Summer School on QCD Perspectives on Hot and Dense Matter}},
  pp.~327--351, 11, 2001
  [\href{https://arxiv.org/abs/hep-ph/0111341}{{\ttfamily hep-ph/0111341}}].

\bibitem{Vanderhaeghen:2000ws}
M.~Vanderhaeghen, J.M.~Friedrich, D.~Lhuillier, D.~Marchand, L.~Van~Hoorebeke
  and J.~Van~de Wiele, \emph{{QED radiative corrections to virtual Compton
  scattering}}, \href{https://doi.org/10.1103/PhysRevC.62.025501}{\emph{Phys.
  Rev. C} {\bfseries 62} (2000) 025501}
  [\href{https://arxiv.org/abs/hep-ph/0001100}{{\ttfamily hep-ph/0001100}}].

\bibitem{Braaten:1989mz}
E.~Braaten and R.D.~Pisarski, \emph{{Soft Amplitudes in Hot Gauge Theories: A
  General Analysis}},
  \href{https://doi.org/10.1016/0550-3213(90)90508-B}{\emph{Nucl. Phys. B}
  {\bfseries 337} (1990) 569}.

\bibitem{Ghiglieri:2020dpq}
J.~Ghiglieri, A.~Kurkela, M.~Strickland and A.~Vuorinen, \emph{{Perturbative
  Thermal QCD: Formalism and Applications}},
  \href{https://doi.org/10.1016/j.physrep.2020.07.004}{\emph{Phys. Rept.}
  {\bfseries 880} (2020) 1} [\href{https://arxiv.org/abs/2002.10188}{{\ttfamily
  2002.10188}}].

\end{thebibliography}\endgroup
\end{document}